\documentclass[a4paper,twocolumn,11pt]{quantumarticle}
\pdfoutput=1
\usepackage[utf8]{inputenc}
\usepackage[english]{babel}
\usepackage[T1]{fontenc}
\usepackage{amsmath}
\usepackage{hyperref}

\usepackage{tikz}
\usepackage{lipsum}

\usepackage{amssymb}
\usepackage{dcolumn}
\usepackage{graphicx}
\usepackage{mathrsfs}
\usepackage{appendix}
\usepackage{graphicx}
\usepackage{booktabs}
\usepackage{color}
\usepackage{bm}
\usepackage{subcaption}
\usepackage{url}
\usepackage{ragged2e}
\usepackage[numbers]{natbib}

\begin{document}

\title{Manipulation of Arbitrary-Order Cavity-Magnon Polariton Blockade}

\author{Zhe-Qi Yang}
\affiliation{Fujian Key Laboratory of Quantum Information and Quantum Optics, College of Physics and Information Engineering, Fuzhou University, Fuzhou, Fujian 350108, China}
\author{Xiao-Yu Bi}
\affiliation{Fujian Key Laboratory of Quantum Information and Quantum Optics, College of Physics and Information Engineering, Fuzhou University, Fuzhou, Fujian 350108, China}
\author{Zhi-Rong Zhong}
\email{zhirz@fzu.edu.cn}
\affiliation{Fujian Key Laboratory of Quantum Information and Quantum Optics, College of Physics and Information Engineering, Fuzhou University, Fuzhou, Fujian 350108, China}
\maketitle

\begin{abstract}
Manipulating the cavity-magnon polariton blockade is significant for achieving precise, on-demand control of individual photons (magnons) and has particular applications in realizing multifunctional quantum technologies, quantum information processing, and hybrid quantum networks. In this paper, we theoretically propose a scheme to realize an $n$-cavity-magnon polariton blockade in a cavity-magnon system by utilizing Kerr nonlinearity. We demonstrate that the Kerr nonlinearity introduces anharmonicity into the polariton energy spectrum, which in turn enables the blockade effect. When the external driving frequency is resonant with the transition to the $n$th polariton excited state, a perfect $n$-polariton blockade is achieved. Moreover, increasing the driving strength can enhance the higher-order blockade while maintaining high fidelity in a dissipative environment. Our work pioneers the field of cavity-magnon polariton blockade, opens a new avenue for the preparation of controllable quantum resources, and holds significant potential for applications in quantum communication and quantum information processing.
\end{abstract}

\section{Introduction}

Photon blockade, arising from the anharmonic energy spectrum induced by optical nonlinearity, suppresses further excitation after a single photon is generated and enables deterministic single-photon sources for quantum information technologies~\cite{PhysRevLett.81.2833}. Conventional photon blockade (CPB) and unconventional photon blockade (UPB)—the latter originating from destructive interference between transition pathways—have been realized in diverse platforms, including cavity quantum electrodynamics (cQED) systems~\cite{Birnbaum2005,PhysRevA.82.032101,doi:10.1126/science.1152261,PhysRevA.97.043819,PhysRevX.7.011012,PhysRevLett.107.053602}, cavity optomechanics~\cite{PhysRevLett.107.063602,PhysRevA.88.023853,PhysRevLett.107.063601,PhysRevA.92.033806,PhysRevA.93.063860}, cavity magnon systems~\cite{https://doi.org/10.1002/qute.202400043,PhysRevA.109.033721,Huang:24} and second-order nonlinear systems \cite{PhysRevB.87.235319,PhysRevA.90.023849,PhysRevA.102.033713}. This concept has recently been extended to multiphoton blockade, where an $n$-photon state is accessible while the $(n+1)$th excitation is suppressed. This advancement offers a new approach to preparing multiphoton quantum states, typically achieved through Kerr nonlinearities \cite{PhysRevA.87.023809,pvws-4ybt,PhysRevLett.121.153601,PhysRevA.110.053702}, photon--atom coupling \cite{PhysRevA.102.053710,Hamsen2018,PhysRevA.110.043707,PhysRevA.107.043702,Lin:24}, or hybrid mechanisms~\cite{PhysRevResearch.6.033247,PhysRevA.104.053718,https://doi.org/10.1002/qute.202300187}. Notably, two-photon blockade has already been experimentally demonstrated in driven cQED systems \cite{PhysRevLett.118.133604}.

Magnons—the quantized quasiparticles of collective spin excitations in ferromagnetic materials such as yttrium iron garnet—have recently emerged as a promising resource for quantum technologies \cite{PhysRevResearch.3.013192,YUAN20221,PhysRevA.108.023501,ASJAD20233,PhysRevB.109.L041301,PhysRevA.103.052419,PhysRevA.103.062605,PhysRevLett.127.183202,PhysRevB.111.064414,zhgm-p3ss,Baghshahi2025}. Analogous to photon blockade, magnon blockade plays a crucial role in generating nonclassical magnon states and single-magnon sources, and both conventional magnon blockade (CMB) \cite{PhysRevA.101.063838,Xu:21,Ding:24} and unconventional magnon blockade (UMB) \cite{PhysRevA.109.043712,Zhang:25,PhysRevA.110.063711} have been theoretically proposed. A recent advancement in this field has extended the blockade effect from individual ``photons'' or ``magnons'' to their hybridized state, the cavity-magnon polariton (CMP) blockade, which has significant applications in quantum information processing, quantum sensing, and nonclassical light sources \cite{PhysRevLett.114.227201,PhysRevLett.104.077202,Zhang2017,Yao2017}. CMPs typically emerge from the strong or even ultrastrong coupling between microwave cavity photons and magnons in yttrium iron garnet (YIG) \cite{10.1063/1.4941730,PhysRevB.93.144420,Zhang2015}, and have been experimentally observed in a microcavity \cite{Kuznetsov2023,Shen2025,Zhang2017}. These features render CMPs a unique quantum resource, stimulating extensive interest and research in a broad range of fields \cite{Rao2019,bnyn-mbwv,Peng:25,Delteil2019,PhysRevApplied.19.014030,PhysRevApplied.23.044065}. However, investigations into polariton blockade have thus far been largely confined to cQED systems and remain restricted to the single-polariton or low-order regimes \cite{nfz3-txyt,PRXQuantum.5.010339}. While single-polariton blockade offers an efficient means of manipulating single quasiparticles for applications in single-particle sources and quantum information processing, its extension to the multi-polariton regime is of profound significance for exploring many-body non-classical states and constructing quantum communication networks. Given that cavity magnon systems are promising candidates for fundamental building blocks in future quantum information infrastructures, it is highly desirable to investigate $n$-CMP blockade ($n$CMPB).

In this work, we propose a scheme for realizing $n$CMPB via the optical Kerr effect. The interaction between the cavity mode and the magnon mode leads to the formation of hybrid polaritonic states, while the Kerr nonlinearity induces anharmonicity in the polariton energy spectrum, thereby enabling the blockade of polariton. In principle, this scheme allows for the realization of $n$CMPB for arbitrary $n$ by tuning the laser detuning, which represents a significant improvement over the work presented in Ref.~\cite{nfz3-txyt}. Furthermore, in a dissipative environment, increasing the external driving strength can effectively enhance the higher-order blockade effect without compromising its fidelity. While single-polariton or low-order blockade schemes are limited to a small set of quantum states and thus restrict applications in quantum information and high-dimensional logic, we demonstrate a controllable blockade of arbitrary $n$-CMPs that enables precise manipulation of many-body nonclassical states. This mechanism provides a feasible route for high-dimensional quantum state engineering and offers a rich, controllable quantum resource for quantum information processing, quantum communication, and quantum sensing. By extending the blockade order, our work lays both experimental and theoretical foundations for high-order quantum logic elements and nonlinear devices, providing a platform for constructing complex quantum networks and promising broad applications in future quantum technologies.

\begin{figure}[t]
    \centering
    \captionsetup[subfigure]{labelformat=empty}
    \begin{subfigure}[t]{0.8\columnwidth}
        \centering
        \includegraphics[width=\textwidth]{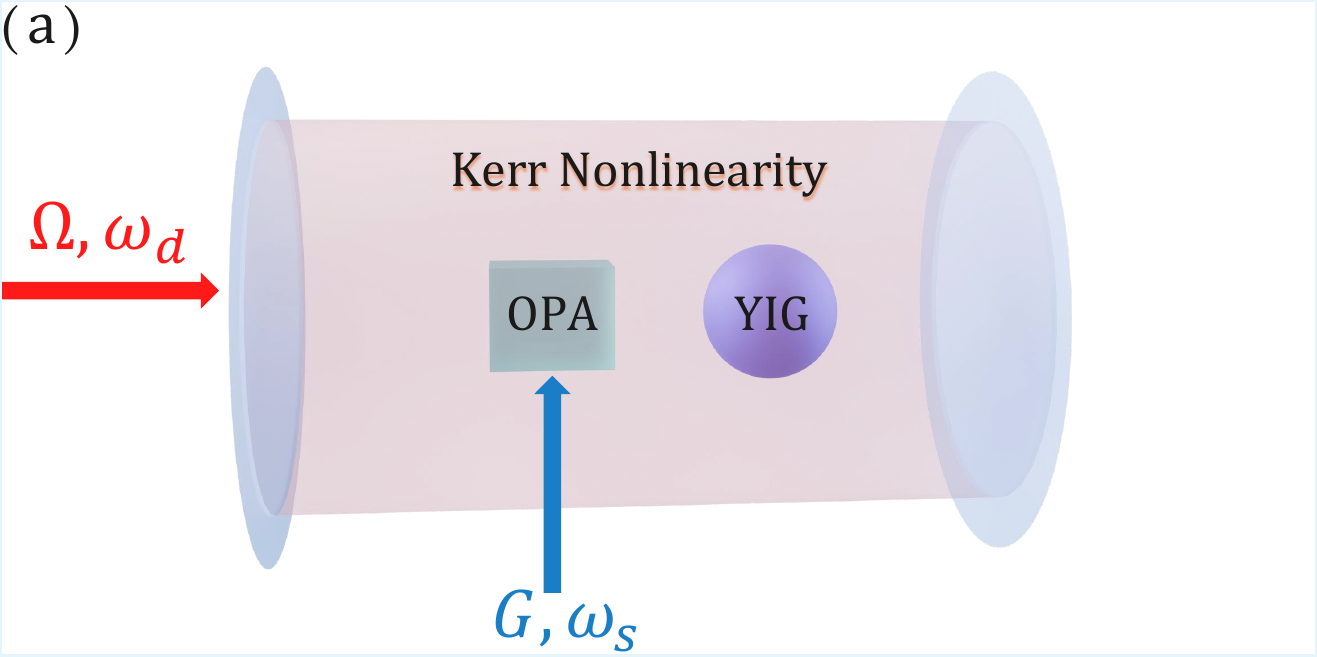}
        \caption{}
        \label{fig:fig1a}

    \end{subfigure}
    \hfill
    \captionsetup[subfigure]{labelformat=empty}
    \begin{subfigure}[t]{0.8\columnwidth}
        \centering
        \includegraphics[width=\textwidth]{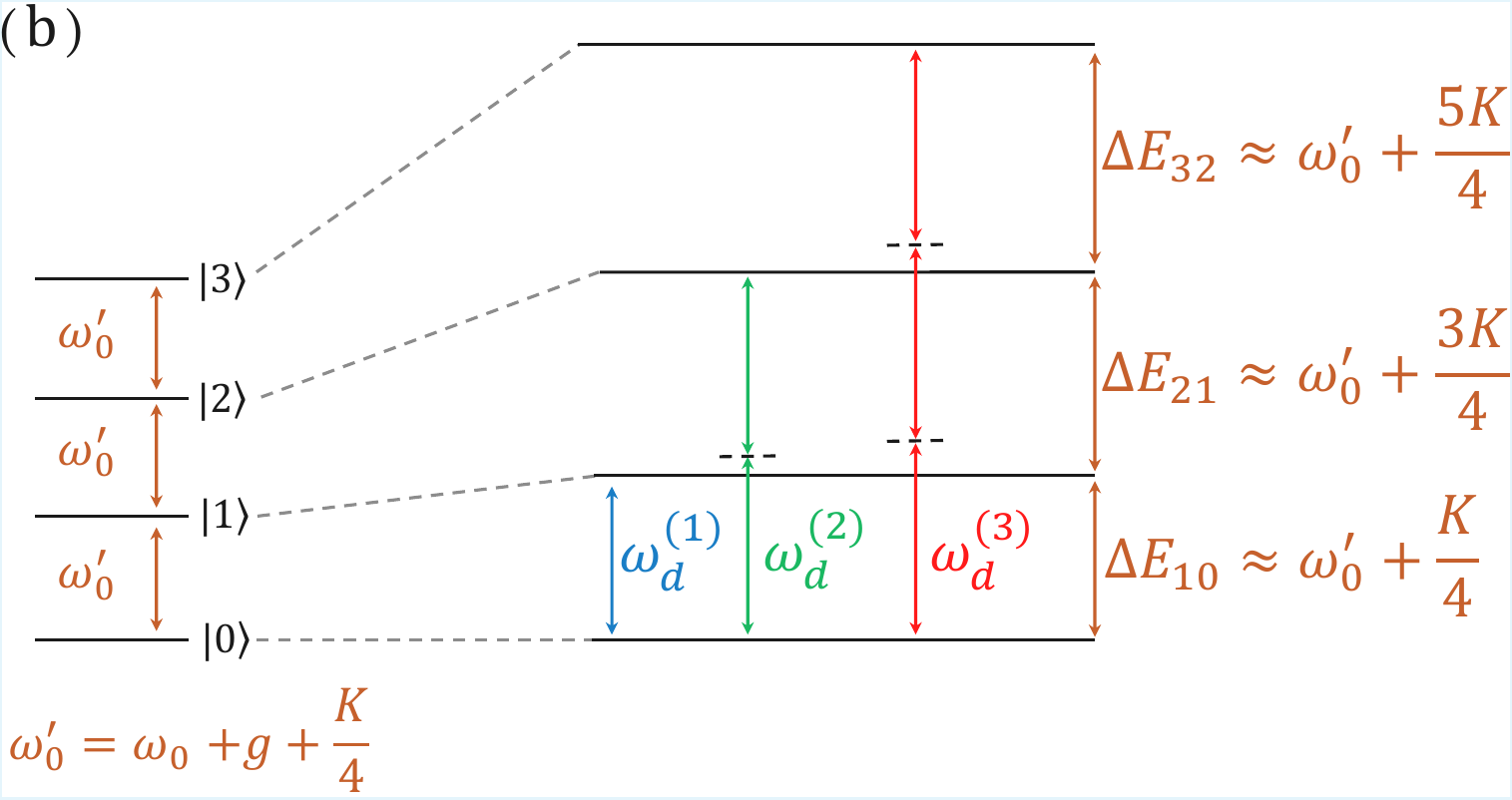}
        \caption{}
        \label{fig:fig1b}

        \end{subfigure}
    \vspace{-8mm}
    \caption{(a) Sketch of the system. (b)Schematic of the $p_+$ polariton energy levels. Here, $\omega_d^{(n)}$ ($n=1, 2, 3$) denotes the external driving frequency required to induce the $n$CMPB, and it satisfies the condition given in Eq. (\ref{eq5}).}

    \label{fig:fig1}
\end{figure}

\section{Theoretical Model}
As illustrated in Fig.~\ref{fig:fig1}\subref{fig:fig1a}, we consider a Kerr nonlinear cavity consisting of a YIG sphere, an optical parametric amplifier (OPA) and a Kerr nonlinear medium. The Hamiltonian of the system can be written as ($\hbar=1$)
\begin{align}
H_{\rm{sys}}&=  \omega_{c}a^{\dagger}a+\omega _{m}m^{\dagger}m + K(a^{\dagger}a)^{2} + g(a^{\dagger}m+am^{\dagger}) \nonumber \\
 &+ \Omega(a^{\dag}e^{-i\omega_{d} t} +ae^{i\omega_{d} t})+G(a^{\dag 2}e^{-i\omega_{s} t} +a^{2}e^{i\omega_{s} t}),\label{eq1}
\end{align}
where $a$ ($a^{\dagger}$) and $m$ ($m^{\dagger}$) represent the annihilation (creation) operators of cavity mode with frequency $\omega_{c}$ and magnon mode with frequency $\omega_{m}$, respectively. $K$ is the Kerr nonlinearity coefficient and $g$ is the cavity-magnon coupling strength. For simplicity, we have absorbed the frequency shift of the cavity mode induced by the Kerr term into the definition of $\omega_c$. The cavity mode is driven by an optical laser with frequency $\omega_{d}$ and amplitude $\Omega$, and interacts with the OPA that is externally pumped by a laser with frequency $\omega_{s}$ and amplitude $G$. In the rotating frame with respect to $\omega_d(a^{\dagger}a+m^{\dagger}m)$, and utilizing $\omega_s=2\omega_d$, the Hamiltonian (\ref{eq1}) can be written as
\begin{align}
H&=H_0+H_d, \nonumber \\
H_0&=  \Delta_{c}a^{\dagger}a+\Delta _{m}m^{\dagger}m + K(a^{\dagger}a)^{2} + g(a^{\dagger}m+am^{\dagger}), \nonumber \\
H_d&= \Omega(a^{\dag} +a)+G(a^{\dag 2} +a^{2}),\label{eq2}
\end{align}
where $\Delta_{c}=\omega_{c}-\omega_{d}$ and $ \Delta_{m}=\omega_{m}-\omega_{d}$. We consider the resonant case where $\Delta_c=\Delta_m=\Delta_0$ ($\omega_c=\omega_m=\omega_0$) and define the polariton modes as $p_{+} = (a + m)/\sqrt{2}$ and $p_{-} = (a - m)/\sqrt{2}$. If we selectively drive the $p_{+}$ mode by setting the driving frequency on resonance with it, the $p_{-}$ mode is then far-detuned and can be safely neglected. In the strong coupling limit ($g\gg K,\Omega, G$), and under the rotating-wave approximation (RWA), Eq. (\ref{eq2}) can be rewritten as
\begin{align}
H_{\rm{eff}}=&\Delta_+ n_{+}+\frac{K}{4}n_{+}^2 + \frac{\Omega}{\sqrt{2}}(p_+^{\dag} +p_+) \nonumber \\
&+\frac{G}{2}(p_+^{\dag 2} +p_+^{2}).  \label{eq4}
\end{align}
where $\Delta_{+}=\Delta_0 + g+\frac{K}{4}$, $n_{+}=p_{+}^{\dagger} p_{+}$. It can be seen from Eq. (\ref{eq4}) that, due to the Kerr effect, namely the presence of the second term on the right-hand side, the energy spacing between adjacent polariton levels is no longer uniform. The energy-level diagram is shown in Fig.~\ref{fig:fig1}\subref{fig:fig1b}. This energy anharmonicity implies that if we set the driving detuning to be on resonance with the $n$th excitation ($\omega_d = \omega_0+g+\frac{K}{4}+ \frac{nK}{4}$), the system will be effectively confined to the subspace with a total excitation number of $N_+ = n$. Any subsequent excitation to the $N_+ = n+1$ subspace will be strongly suppressed, as the drive is now far off-resonance from that transition frequency. Therefore, the condition for the $n$CMPB can be simply written as
\begin{align}
\Delta_+ =  -\frac{nK}{4}, n=1,2,3,\cdots .\label{eq5}
\end{align}
It is important to note that this condition can also be derived directly from the original full Hamiltonian in Eq.~(\ref{eq1}), even without performing the polariton transformation. We provide a detailed derivation in Appendix \hyperlink{app:A}{A}, where we also demonstrate the validity of the effective Hamiltonian in Eq. (\ref{eq4}). Notably, we analyze the blockade effect by examining the evolution of the system's population and the average particle number \cite{PhysRevA.87.023809}. This approach allows a direct and effective characterization of the blockade, regardless of whether dissipation is present.

\begin{figure}[t]
    \centering
    \captionsetup[subfigure]{labelformat=empty}
    \begin{subfigure}[t]{0.8\columnwidth}
        \centering
        \includegraphics[width=\textwidth]{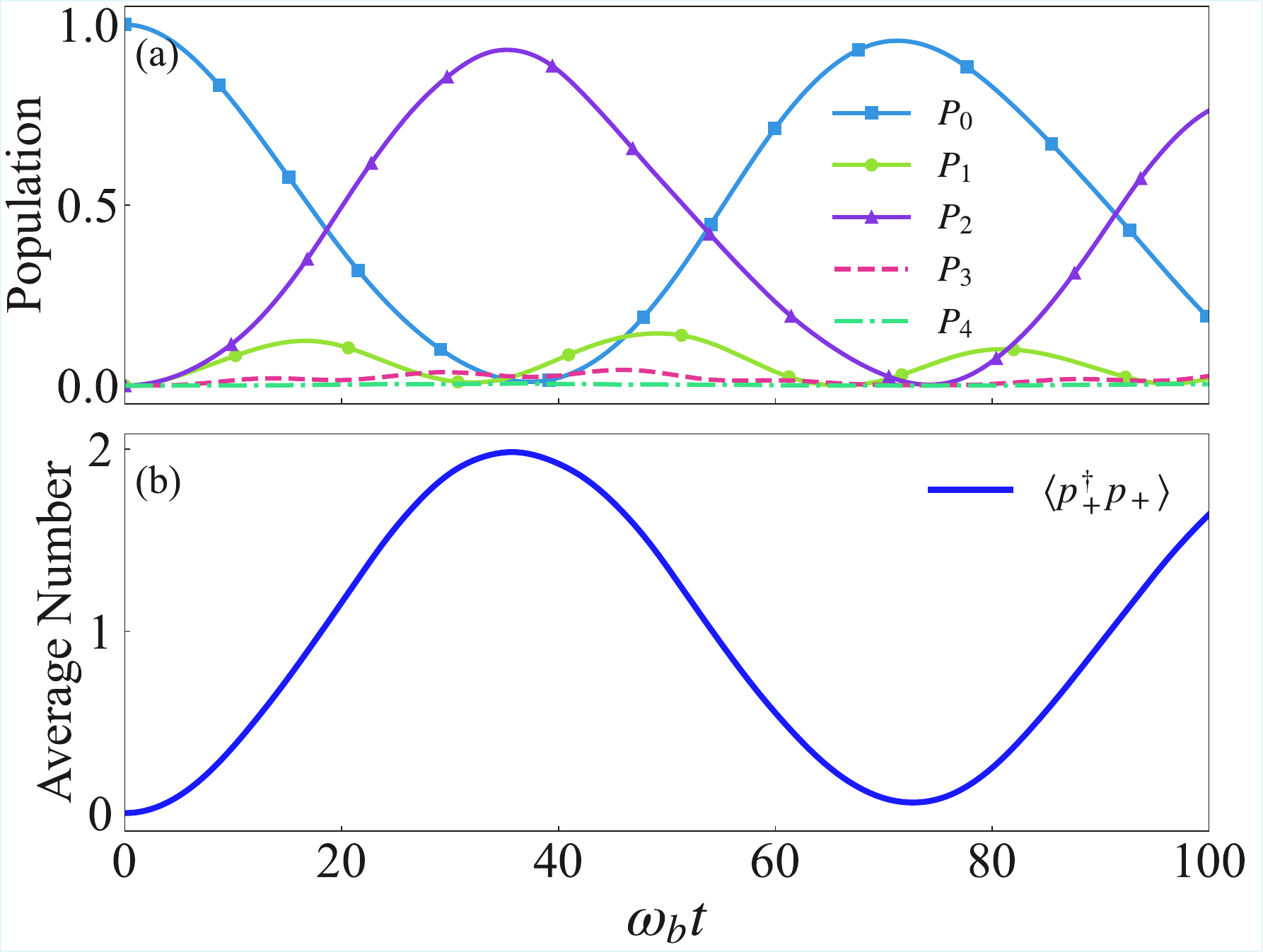}
        \caption{}
        \label{fig:fig2a}
    \end{subfigure}
    \vspace{-5mm}
    \captionsetup[subfigure]{labelformat=empty}

    \begin{subfigure}[t]{0.8\columnwidth}
        \centering
        \includegraphics[width=\textwidth]{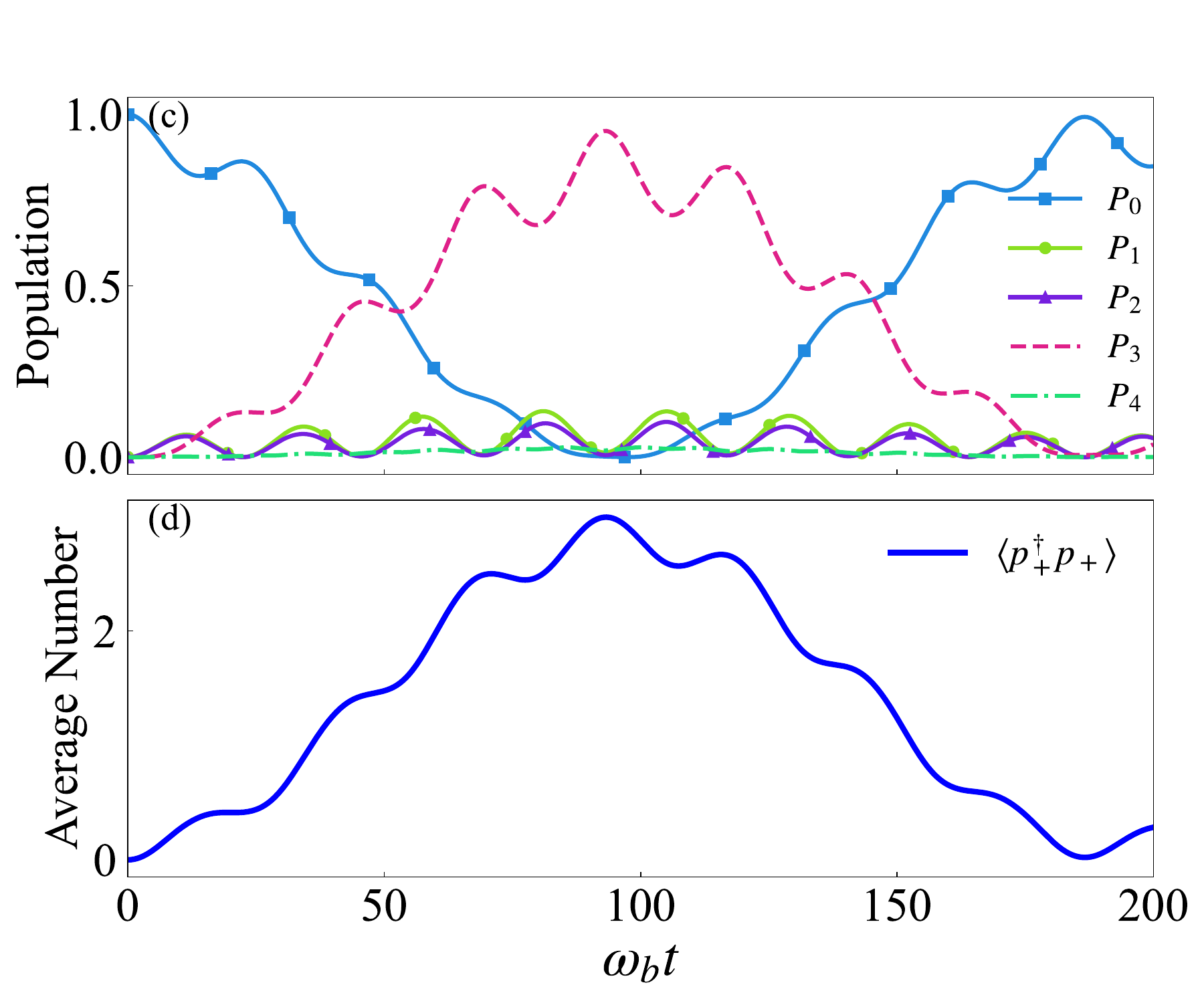}
        \caption{}
        \label{fig:fig2b}
        \end{subfigure}

    \begin{subfigure}[t]{0pt}
        \caption{}
        \label{fig:fig2c}
    \end{subfigure}

     \begin{subfigure}[t]{0pt}
        \caption{}
        \label{fig:fig2d}
    \end{subfigure}
    \vspace{-20mm}
    \caption{Time evolution of the excitation number distribution for the (a) $n$ = 2 and (c) $n$ = 3 cases. $P_{j}= \langle \psi(t)|j \rangle \langle j | \psi(t) \rangle $ ($j=0,1,2,3,4$). Here, $|j \rangle \langle j|$ is the projection operator onto the $j$-th Fock state of the $p_+$ mode. Time evolution of the mean excitation number for the (b) $n$ = 2 and (d) $n$ = 3. The average particle number $\langle p_{+}^{\dagger} p_{+} \rangle =  \langle \psi(t) |p_{+}^{\dagger} p_{+}| \psi(t) \rangle $. $\Delta_+/g = -0.025$ in panels (a) and (b) and $\Delta_+/g = -0.0375$ in panels (c) and (d). System parameters are $K/g=0.05$, and $\Omega/g=G/g=0.005$.}
    \label{fig:fig2}
\end{figure}

\section{$N$-CMP Blockade}
Figures~\ref{fig:fig2}\subref{fig:fig2a} and~\ref{fig:fig2}\subref{fig:fig2b}, as well as Figs.~\ref{fig:fig2}\subref{fig:fig2c} and~\ref{fig:fig2}\subref{fig:fig2d}, respectively show, in the absence of dissipation, the excitation-number distributions and the time evolution of the mean excitation numbers for $n$ = 2 and $n$ = 3. As seen in Fig.~\ref{fig:fig2}\subref{fig:fig2a}, the excitation number distribution is heavily suppressed for $n > 2$. Throughout the evolution, $P_2$ peaks at a value of 0.93 while $P_3$ remains suppressed at a maximum of just 0.04, which is a clear manifestation of the 2CMPB. In this scenario, the system's dynamics can be approximated as a Rabi oscillation between the ground state and the state $|2 \rangle$. For the $n$ = 3 case, the blockade effect is even more pronounced: the maximum value of $P_3$ reaches 0.95, while $P_4$ is approximately zero, as shown in Fig.~\ref{fig:fig2}\subref{fig:fig2c}. This phenomenon is readily understood: for higher energy levels, the anharmonic energy shift between adjacent levels increases. This larger detuning from the driving frequency results in a stronger and more effective blockade.

\begin{figure}[t]
    \centering
    \captionsetup[subfigure]{labelformat=empty}

    \begin{subfigure}[t]{0.49\columnwidth}
        \centering
        \includegraphics[width=\textwidth]{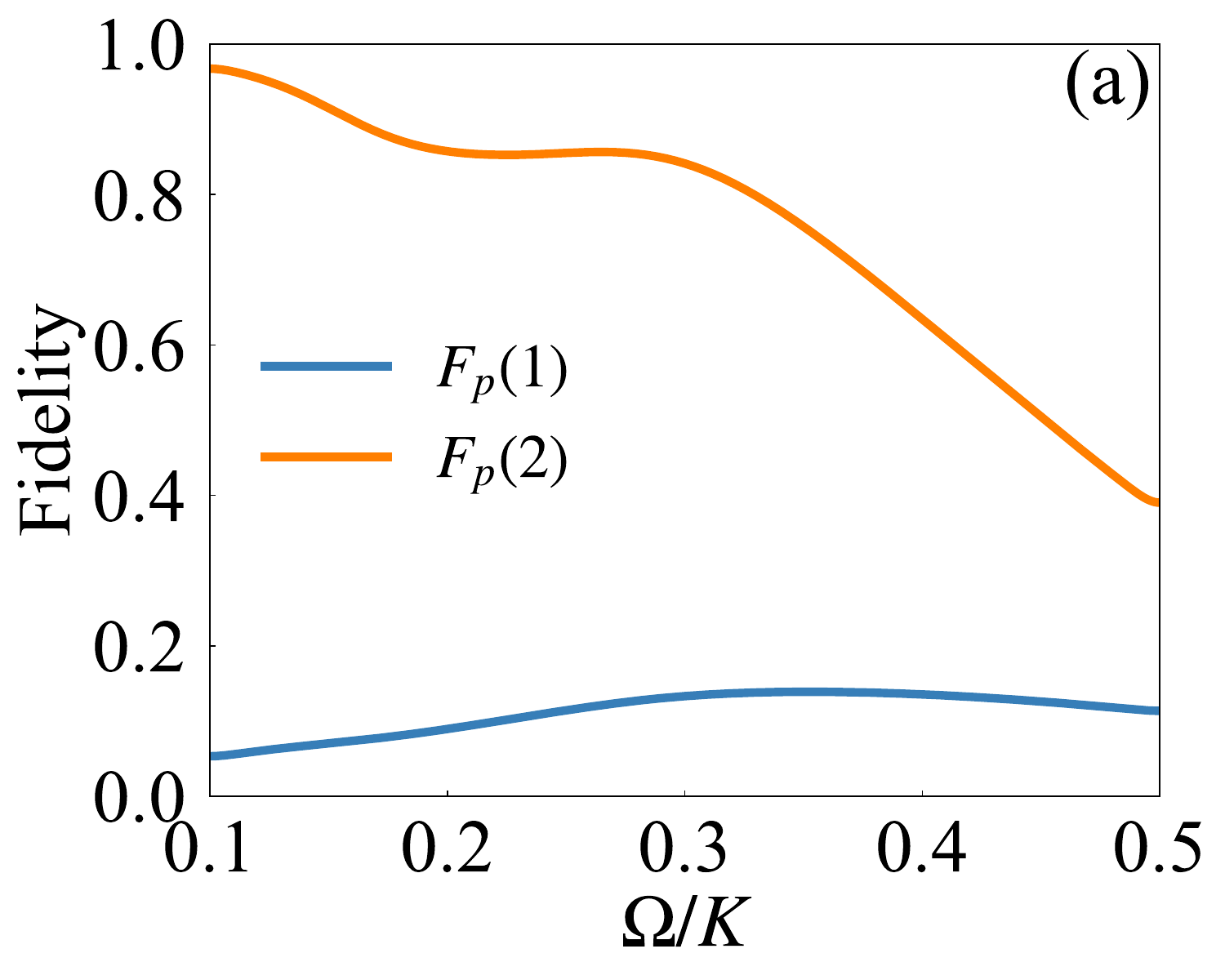}
        \caption{}
        \label{fig:fig3a}
    \end{subfigure}
    \hfill
    \begin{subfigure}[t]{0.49\columnwidth}
        \centering
        \includegraphics[width=\textwidth]{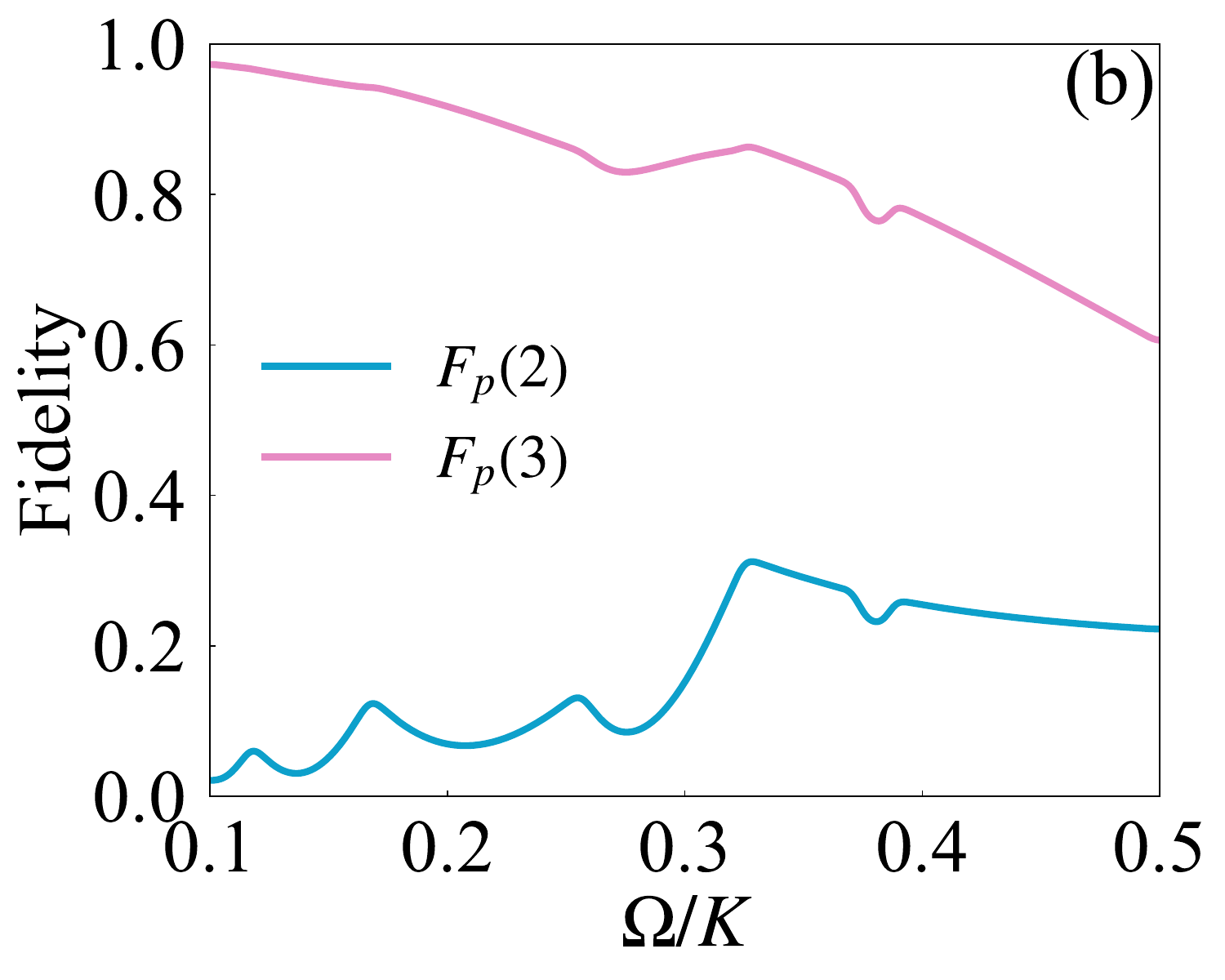}
        \caption{}
        \label{fig:fig3b}
    \end{subfigure}
    \vspace{-5mm}

    \caption{Fidelity $F_p(n)$ as a function of $\Omega/K$ for the cases of (a) $n$ = 2 and (b) $n$ = 3. $\Delta_+/g =-0.025$ in panel (a) and $\Delta_+/g = -0.0375$ in panel (b). System parameters are $K/g=0.05$ and $G/g=0.005$.}
    \label{fig:fig3}
\end{figure}

To achieve blockade, the system must also be in the weak-driving regime, which requires the condition $\Omega,G \ll K$ \cite{PhysRevA.87.023809}. We define the fidelity of the $n$CMPB as
\begin{align}
F_p(n) = \sum_{j=0}^{n}	\langle \psi(t)| j \rangle \langle j|  \psi(t) \rangle .  \label{eq6}
\end{align}
We consider the $n$CMPB to be successfully realized if two conditions are met: (i) the fidelity $F_p(n)\approx 1$, and (ii) for $j < n$, the fidelity $F_p(j) \ll 1$. If the weak-driving condition is not satisfied, the energy anharmonicity induced by the Kerr effect will be insufficient to prevent the drive from exciting the system to higher energy levels. This leads to a degradation of the fidelity $F_p(n)$. Therefore, in Fig.~\ref{fig:fig3}, we have plotted the fidelity $F_p$ as a function of the driving strength $\Omega$, for the $n$ = 2 and $n$ = 3 cases. As shown in Fig.~\ref{fig:fig3}\subref{fig:fig3a}, at the beginning of the scanned range ($\Omega/K$ = 0.1), the fidelity $F_p(2)$ is found to be approximately unity, while the fidelity $F_p(1)$ is significantly less than 1. This confirms that the system is successfully blockaded in the total excitation subspace of $N = 2$. As $\Omega/K$ increases, the aforementioned fidelities exhibit a general downward trend. This indicates a failure to maintain an effective 2CMPB, as the strong drive begins to excite the system into subspaces with a total excitation number $N > 2$. The situation for $n$ = 3, shown in Fig.~\ref{fig:fig3}\subref{fig:fig3b}, is qualitatively similar. However, a key difference is that the fidelity for $n$ = 3 exhibit a less pronounced decline with increasing driving strength compared to the $n$ = 2 case. This is because a stronger drive is required to overcome the energy anharmonicity and excite the system into the $N > 3$ subspace. More significant fluctuations are observed in $F_p(2)$ in Fig.~\ref{fig:fig3}\subref{fig:fig3b} relative to $F_p(1)$ in Fig.~\ref{fig:fig3}\subref{fig:fig3a}. This can be attributed to the increased population of the $|4 \rangle$ state at higher driving strengths. Transitions between the $|2 \rangle$ and $|4 \rangle$ states are subsequently induced by the two-photon drive, resulting in larger oscillations in the $F_p(2)$ fidelity.

\begin{figure}[t]
    \centering
    \captionsetup[subfigure]{labelformat=empty}
    \begin{subfigure}[t]{0.8\columnwidth}
        \centering
        \includegraphics[width=\textwidth]{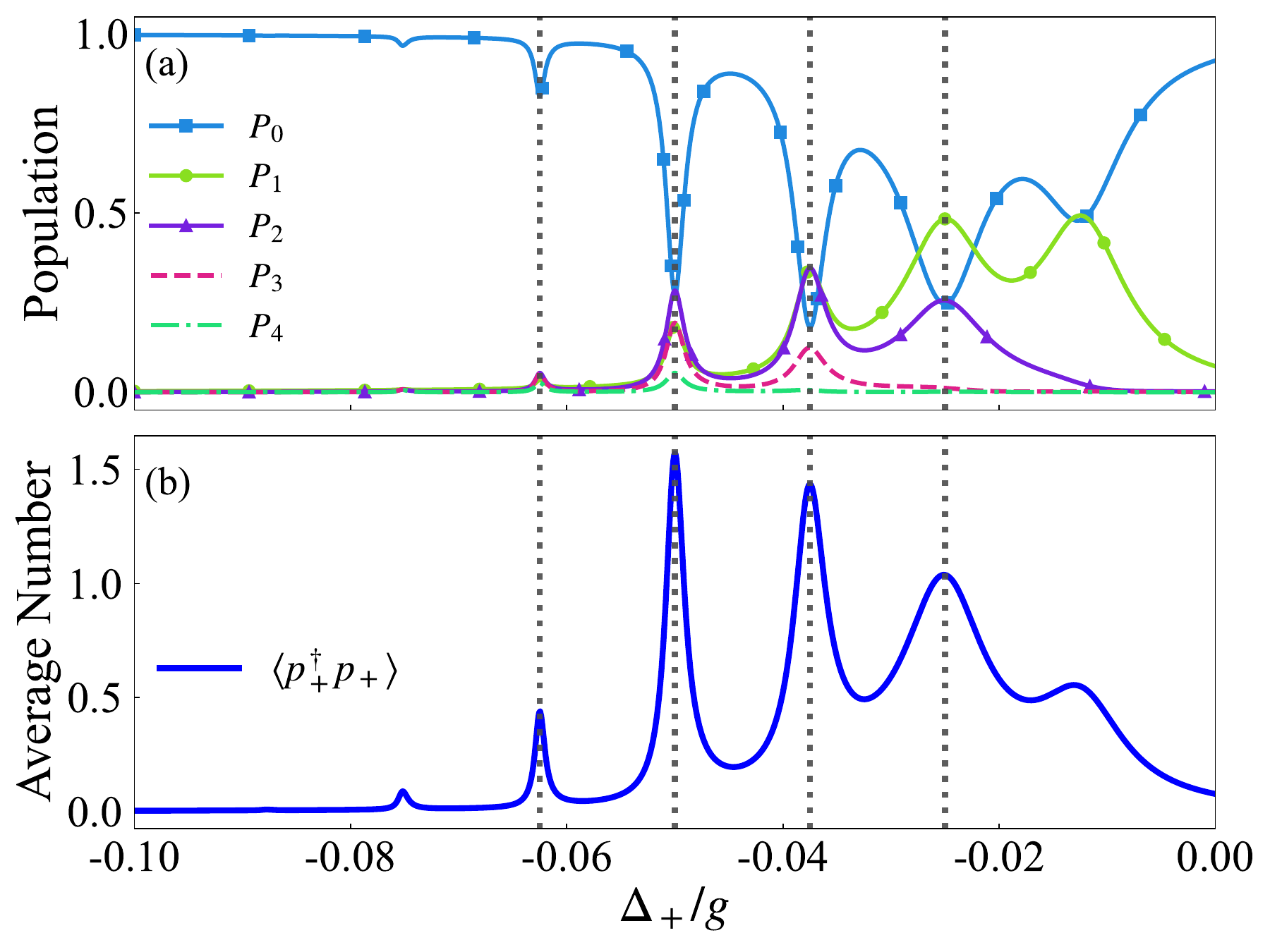}
        \caption{}
        \label{fig:fig4a}
    \end{subfigure}
    \vspace{-6mm}
    \captionsetup[subfigure]{labelformat=empty}

    \begin{subfigure}[t]{0.8\columnwidth}
        \centering
        \includegraphics[width=\textwidth]{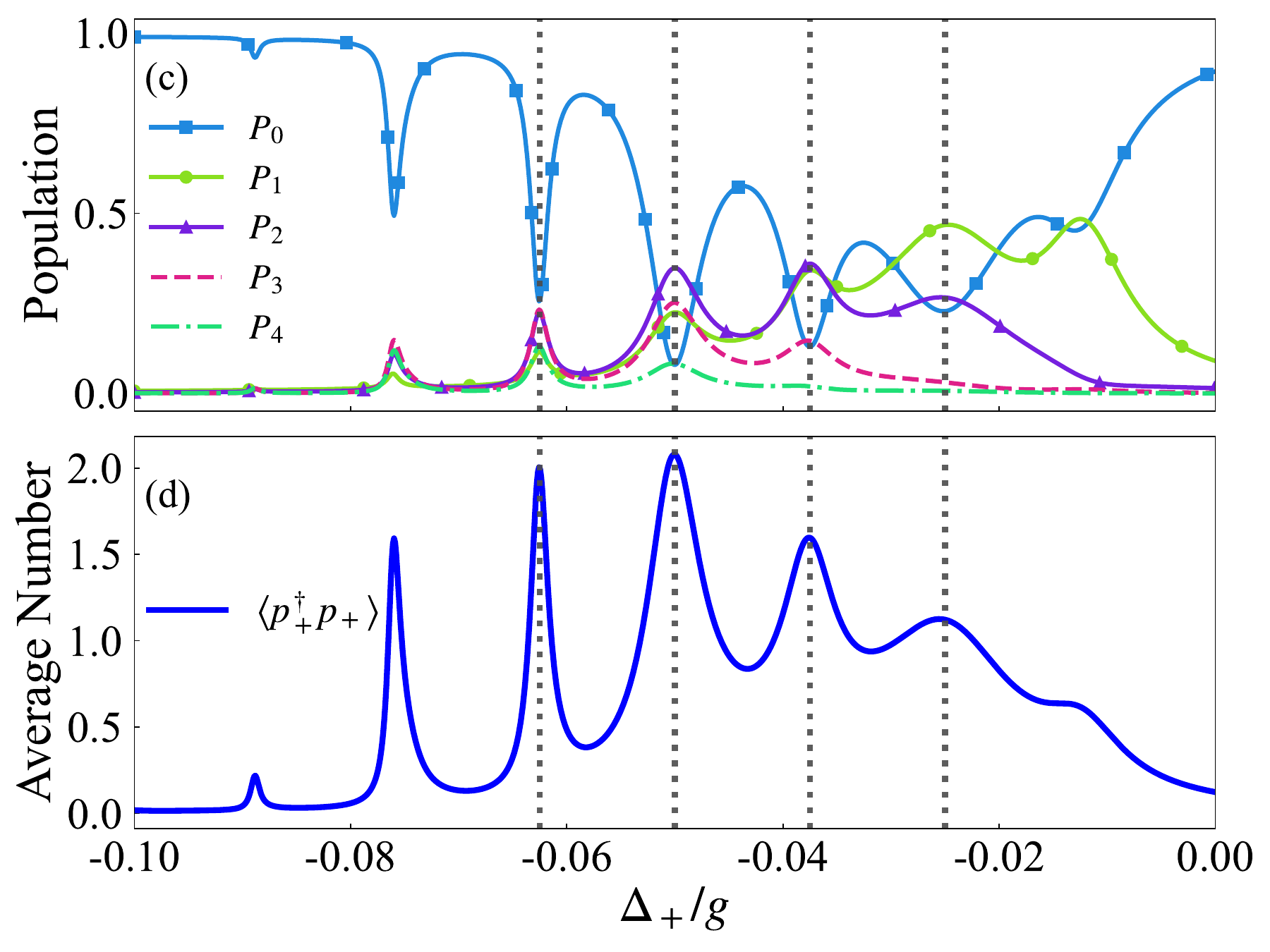}
        \caption{}
        \label{fig:fig4b}
        \end{subfigure}

    \begin{subfigure}[t]{0pt}
        \caption{}
        \label{fig:fig4c}
    \end{subfigure}

     \begin{subfigure}[t]{0pt}
        \caption{}
        \label{fig:fig4d}
    \end{subfigure}
    \vspace{-20mm}
    \caption{Excitation number probability distribution, and mean excitation number as a function of the detuning, $\Delta_+/g$, in the steady state $\rho_{ss}$. $P_{j}= \mathrm{Tr}(| j \rangle \langle j|\rho_{ss})  $ ($j=0,1,2,3,4$). $\langle p_{+}^{\dagger} p_{+} \rangle =  \mathrm{Tr}(p_{+}^{\dagger} p_{+}\rho_{ss})$. $G/g=0.005$ in panels (a) and (b) and $G/g=0.01$ in panels (c) and (d). The gray dashed lines in the figure (from right to left) indicate the theoretical conditions for the $n$PB for $n$ = 2, 3, 4, and 5, respectively. System parameters are $K/g=0.05$, $\Omega/g=0.005$ and $\kappa/g=0.001$.}
    \label{fig:fig4}
\end{figure}

So far, we discussed the eigenstates and eigenvalues of $H_0$, established the conditions for blockade, and analyzed the corresponding evolution of the closed system under these conditions. We now extend our analysis to an open system and investigate the effects of dissipation on the dynamics. We consider the dynamics of the system to be governed by the Lindblad master equation $\dot{\rho}(t)=-i[H_{\rm{eff}}, \rho] + \kappa \mathcal{L}[p_+] \rho$, where $\rho$ is the system density operator, $\kappa$ is the decay rates of the polariton modes, respectively. The Lindblad superoperator is defined as $\mathcal{L}[p_+]\rho = (2p_+\rho p_+^\dagger - \rho p_+^\dagger p_+ - p_+^\dagger p_+ \rho)/2$ and $\rho_{ss}$ denotes the steady-state solution to the master equation in the limit of $t \to \infty$.

The first question to address is how the system's evolution is modified by the inclusion of dissipation. Figure~\ref{fig:fig4} shows the steady-state excitation number probability distribution, as well as the dependence of the excitation number on the $\Delta_+$. The plotted quantities are calculated from $\rho_{ss}$. By analyzing Figs.~\ref{fig:fig4}\subref{fig:fig4a} and~\ref{fig:fig4}\subref{fig:fig4b}, we can identify the conditions for various types of polariton blockade effects. For instance, the first and second dashed lines from the right in the figure correspond to the conditions for the 2CMPB and 3CMPB, respectively. The detuning values $\Delta_+$ at these lines are in excellent agreement with the values selected for the analysis in Fig.~\ref{fig:fig2}. While the signatures of higher-order blockade effects are still observable in the figure, the occupation probabilities of the corresponding high-number Fock states are significantly suppressed. For example, in the 4CMPB and 5CMPB cases, although the probabilities of populating the states $| 1 \rangle$, $| 2 \rangle$, $| 3 \rangle$, and $| 4 \rangle$ are non-zero, they are dwarfed by the probability of the system remaining in the vacuum state, $| 0 \rangle$. Therefore, to achieve a high-quality, higher-order blockade, we increase the strength of the two-photon drive, $G$. As shown in Figs.~\ref{fig:fig4}\subref{fig:fig4c} and~\ref{fig:fig4}\subref{fig:fig4d}, under the conditions for higher-order blockade, the mean excitation number is significantly enhanced compared to the case with the previously weaker drive. Concurrently, the population of the ground state is reduced, indicating that a more significant fraction of the system's population has been transferred to higher number states. For the 5CMPB case, the mean excitation number increases from approximately 0.5 to 2.0, and the occupation probability of the excited state becomes comparable to that of the ground state. This represents a substantial improvement in achieving a high-quality, higher-order blockade. Furthermore, a signature of a 6CMPB emerges in Figs.~\ref{fig:fig4}\subref{fig:fig4c} and~\ref{fig:fig4}\subref{fig:fig4d} at $\Delta_+/g = -0.075$. Although the ground state remains the most populated state in this regime, the population of the higher excited states is still a significant enhancement over the results shown in Figs.~\ref{fig:fig4}\subref{fig:fig4a} and~\ref{fig:fig4}\subref{fig:fig4b}. However, this approach has a potential pitfall: the strong drive may overcome the blockade and excite the system to Fock states beyond the intended target.

\begin{figure}[t]
    \centering
    \captionsetup[subfigure]{labelformat=empty}

    \begin{subfigure}[t]{0.49\columnwidth}
        \centering
        \includegraphics[width=\textwidth]{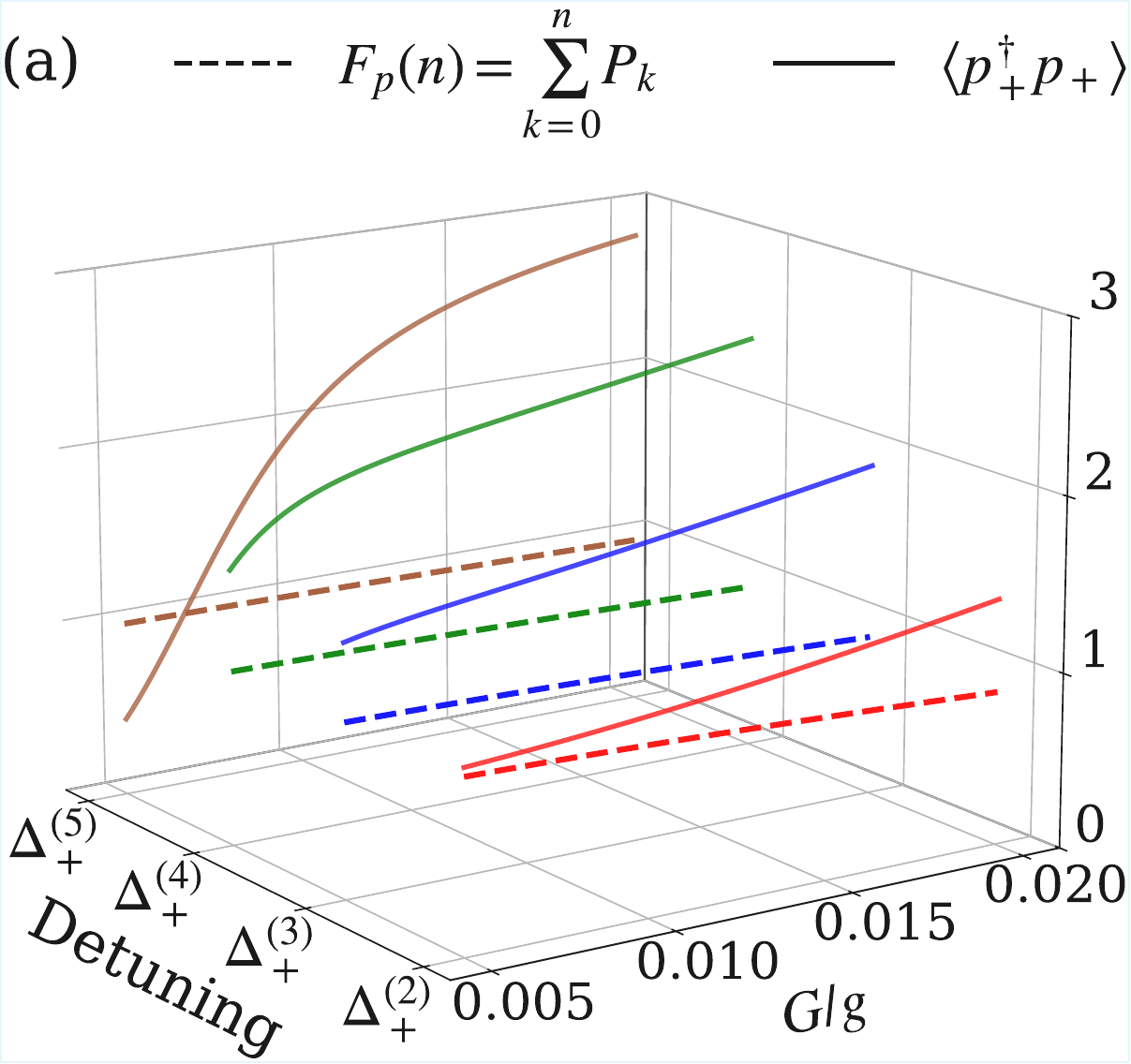}
        \caption{}
        \label{fig:fig5a}
    \end{subfigure}
    \hfill
    \begin{subfigure}[t]{0.49\columnwidth}
        \centering
        \includegraphics[width=\textwidth]{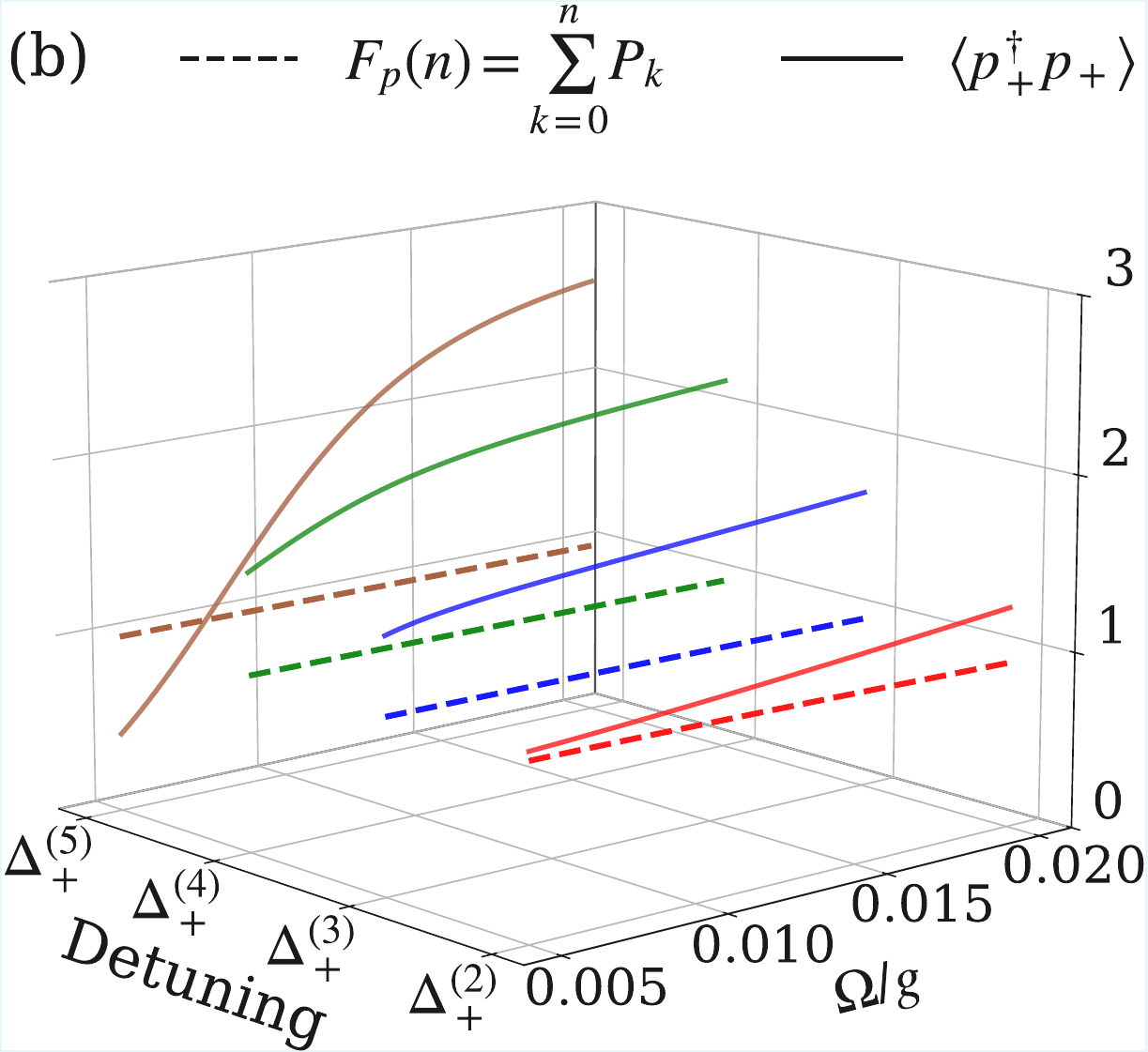}
        \caption{}
        \label{fig:fig5b}
    \end{subfigure}
    \vspace{-5mm}
    \caption{Fidelities $F_p(n)$ and the mean excitation number as a function of the drive strength (a) $G$ and (b) $\Omega$ for different $n$CMPB conditions. The four sets of colored lines in the figure represent the scenarios for different $n$CMPBs. For instance, the red dashed and solid lines respectively depict the variation of $F_p(2)$ and $\langle p_{+}^{\dagger} p_{+} \rangle$ with $G/g$ under the condition of 2CMPB. System parameters are $K/g=0.05$ and $\kappa/g=0.001$. $\Omega/g=0.005$ in panel (a) and $G/g=0.005$ in panel (b).}
    \label{fig:fig5}
\end{figure}

To investigate the influence of the drive strengths, $\Omega$ and $G$, on the blockade effect in the presence of dissipation, we have plotted two key metrics in Fig.~\ref{fig:fig5}. Specifically, for different types of blockade, the figure shows the fidelity and the mean excitation number as a function of the corresponding drive strength. The y-axis label, $\Delta_+^{(n)}$, denotes the value of the detuning $\Delta_+$ required to achieve the $n$CMPB. The values for $n$ = 2, 3, 4, and 5 correspond sequentially to the four gray dashed lines shown from right to left in Fig.~\ref{fig:fig4}. As shown in Fig.~\ref{fig:fig5}\subref{fig:fig5a}, for the higher-order blockade effects, increasing the drive strength $G$ leads to a significant enhancement in the mean excitation number, while the fidelity is maintained at a high level, $F_a(n) \approx 1$. For the 5CMPB case, the mean excitation number is substantially increased from 0.44 to 2.82, while the fidelity decreases by only a negligible amount, from approximately 1.0 to 0.97. This indicates that increasing the two-photon drive strength can effectively populate higher number states without compromising the integrity of the blockade effect. It is worth noting, however, that the situation for the 2CMPB is markedly different. In this case, the mean excitation number shows no significant increase, while in contrast, the fidelity undergoes a substantial degradation, dropping from 0.99 to 0.87. A similar effect is observed for the single-photon drive, $\Omega$. In contrast to the two-photon drive, increasing the single-photon drive strength is less effective at enhancing the mean excitation number, as illustrated in Fig.~\ref{fig:fig5}\subref{fig:fig5b}. However, the corresponding impact on fidelity is significantly less detrimental. For instance, for the 2CMPB, the fidelity only decreases from 0.99 to 0.94, while for the 5CMPB, it is maintained at a near-unity level of 1.0. This conclusion is in apparent contradiction with our earlier discussion of Fig.~\ref{fig:fig3}, where we required the weak-driving condition $\Omega,G \ll K$. However, that requirement applies strictly to the ideal, dissipationless case, which was the focus of our analysis for Fig.~\ref{fig:fig3}. In a dissipative system, dissipation itself contributes to the blockade effect. This is because a larger dissipation rate makes it more difficult to populate higher-lying excited states, as they decay rapidly. Consequently, the weak-driving condition can be relaxed in the presence of dissipation. Therefore, by judiciously tuning the drive strengths $G$ and $\Omega$, we can optimize the performance of a targeted polariton blockade. Additionally, by tuning the system parameters, we can selectively achieve single-photon or single-magnon blockade. A detailed discussion is provided in Appendix \hyperlink{app:B}{B}. Our system thus exhibits high tunability and flexibility.

\section{Discussion and Conclusions}
Our proposal is based on a standard cavity magnon system. The frequencies of the cavity and magnon modes are chosen to be 10.1 GHz, a value typically adopted in both theoretical and experimental studies \cite{PhysRevLett.120.057202,ZARERAMESHTI20221,PhysRevApplied.12.034001,PhysRevResearch.1.023021,Kong:21}. Typically, the cavity-magnon coupling strength $g$ ranges from 1 to 100 MHz \cite{Shen2025,10.1063/5.0006753,PhysRevLett.133.043601}, but it can reach the GHz scale upon entering the ultrastrong coupling regime \cite{RevModPhys.91.025005,PhysRevApplied.2.054002,Zhang2025}. At present, achieving strong Kerr nonlinearity in optical cavities (e.g., with a Kerr coefficient exceeding the cavity dissipation rate) remains highly challenging. Nevertheless, strong Kerr nonlinearities have been experimentally realized in superconducting microwave cavities \cite{Grimm2020,He2023,Ding2025,Iyama2024}, and a number of theoretical works have also adopted relatively strong Kerr nonlinearities in numerical simulations~\cite{Ding:24,PhysRevA.87.023809,PhysRevA.104.053718,pvws-4ybt,PhysRevA.109.033721,Xu:25,PhysRevResearch.6.033247}. For instance, in Ref.~\cite{Iyama2024}, a Kerr parametric oscillator (KPO) was utilized to realize the Kerr effect, achieving a Kerr coefficient of $K/2\pi = 3.1$ MHz. Ref.~\cite{He2023} employs a superconducting nonlinear asymmetric inductive element (SNAIL)-terminated resonator coupled to an ancillary qubit to realize tunable Kerr nonlinearity, with the Kerr coefficient adjustable from near zero to about $K/2\pi \approx 5-6$ MHz. Therefore, the parameters employed in our numerical simulations are well within the experimentally accessible range.

In conclusion, we have proposed a scheme for realizing arbitrary $n$-cavity-magnon polariton blockade. By introducing Kerr nonlinearity into the system, anharmonicity is induced in the adjacent polariton energy levels, thereby enabling the polariton blockade effect. Furthermore, by analyzing the system dynamics in a dissipative environment, we find that increasing the driving strength can significantly enhance the higher-order blockade without compromising its fidelity. Our work transcends the limitations of previous single-polariton or low-order blockade and pioneers the field of cavity-magnon polariton blockade, providing a viable route for the precise preparation of many-body non-classical quantum states. This holds significant potential for various applications, including quantum information processing, quantum computing, and quantum sensing.

\section*{Author Contributions}
Z.-Q. Yang performed the numerical simulations and wrote the manuscript. X.-Y. Bi contributed to the manuscript preparation and assisted with the numerical code development. Z.-R. Zhong conceived and supervised the project, and revised the manuscript. All authors discussed the results and approved the final version of the manuscript.

\section*{Acknowledgements}
This work was supported by the National Natural Science Foundation of China (12074070).

\onecolumn
\appendix
\hypertarget{app:A}{}
\section{Validity of The Effective Hamiltionian}
\begin{figure}[t]
    \centering
    \captionsetup[subfigure]{labelformat=empty}
    \begin{subfigure}[t]{0.6\columnwidth}
        \centering
        \includegraphics[width=\textwidth]{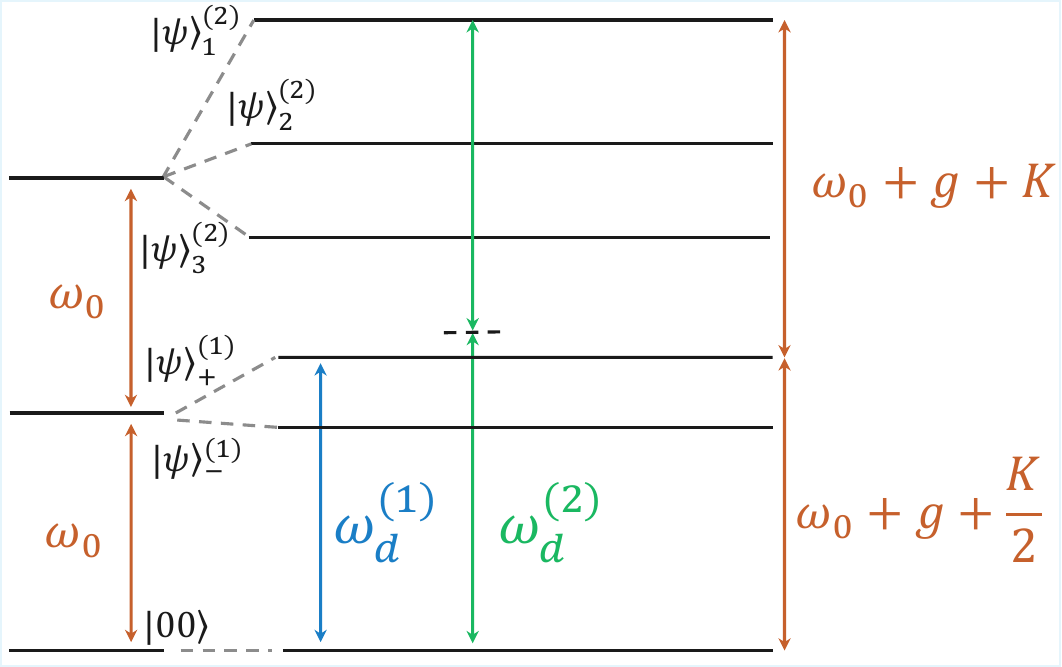}
        \caption{}
        \label{fig:figS1a}

    \end{subfigure}

    \caption{Energy level of the system. Due to the strong cavity-magnon coupling, the energy levels undergo a splitting. The upper and lower energy levels correspond to the $p_+$ and $p_-$ polariton modes, respectively.}
    \label{fig:figS1}
\end{figure}
In the main text, we performed a polariton transformation on the original Hamiltonian (\ref{eq1}) to obtain an effective Hamiltonian. From this, we derived the condition for the $n$CMPB, given by Eq. (\ref{eq5}). In fact, this condition can also be derived directly from the original Hamiltonian without performing the polariton transformation. We start from the Hamiltonian in Eq. (\ref{eq1}), which for clarity, we reproduce here ($\hbar=1$):
\begin{align}
H^{'}&=H_0^{'}+H_d^{'} \nonumber \\
H_0^{'}&=  \omega_{0}a^{\dagger}a+\omega _{0}m^{\dagger}m + K(a^{\dagger}a)^{2} + g(a^{\dagger}m+am^{\dagger}) \nonumber \\
H_d^{'}&= \Omega(a^{\dag}e^{-i\omega_{d} t} +ae^{i\omega_{d} t})+G(a^{\dag 2}e^{-i\omega_{s} t} +a^{2}e^{i\omega_{s} t}).\label{eqs1}
\end{align}
If we perform a polaron transformation on Eq.~(\ref{eqs1}), it can be written as:
\begin{align}
H = & \left( \frac{\Delta_c + \Delta_m}{2} + g \right) p_+^\dagger p_+ 
   + \left( \frac{\Delta_c + \Delta_m}{2} - g \right) p_-^\dagger p_- 
   + \frac{\Delta_c - \Delta_m}{2} (p_+^\dagger p_- + p_-^\dagger p_+) \nonumber \\
   & + \frac{K}{4} \left( p_+^\dagger p_+ + p_-^\dagger p_- + p_+^\dagger p_- + p_-^\dagger p_+ \right)^2 + \frac{\Omega}{\sqrt{2}} (p_+^\dagger + p_+ + p_-^\dagger + p_-) \nonumber \\
   & + \frac{G}{2} \left( p_+^{\dagger 2} + p_-^{\dagger 2} + p_+^2 + p_-^2 + 2p_+^\dagger p_-^\dagger + 2p_+ p_- \right),\label{eqs2}
\end{align}
where $\Delta_{c}=\omega_{c}-\omega_{d}$, $ \Delta_{m}=\omega_{m}-\omega_{d}$ and $p_{\pm} = (a \pm m)/\sqrt{2}$. By setting $\Delta_c=\Delta_m=\Delta_0$ and driving solely the $p_+$ mode, i.e., $\Delta_0+g-\frac{K}{4} =  -\frac{nK}{4}$, $n=1,2,3,\cdots$, and under the strong coupling condition $g \gg \{ K,\Omega, G \}$, Eq.~(\ref{eqs2}) can be written, within the rotating-wave approximation, as
\begin{align}
H_{\rm{eff}}&=\Delta_+ n_{+}+\frac{K}{4}n_{+}^2 + \frac{\Omega}{\sqrt{2}}(p_+^{\dag} +p_+)+\frac{G}{2}(p_+^{\dag 2} +p_+^{2}),\label{eqs0}
\end{align}
where $\Delta_{+}=\Delta_0 + g +\frac{K}{4}$ and $n_{\pm}=p_{\pm}^{\dagger} p_{\pm}$. Since the $p_-$ mode oscillates as $p_- \sim e^{i2gt}$, in the strong coupling regime all terms involving $p_-$ can be safely neglected. Equation~(\ref{eqs0}) here actually corresponds to Eq.~(\ref{eq4}) in the main text. In the main text, we focus on driving only the $p_+$ mode. We note that the total excitation number, $N=a^{\dagger}a+m^{\dagger}m$, is conserved by $H_0^{'}$. In principle, this allows us to find the eigenstates and eigenvalues of $H_0^{'}$ for any given $N$. We now consider the specific cases of $N = 1$ and $N = 2$ as illustrative examples and provide their corresponding eigenvalues and eigenstates. For the $N = 1$ case, the eigenvalue and eigenstate of $H_0^{'}$ can be readily determined as
\begin{align}
\psi_{\pm}^{(1)}&=\cos \theta\left|1,0\right\rangle \pm \sin \theta\left|0,1\right\rangle,\label{eqs3}
\end{align}
and
\begin{align}
E_{\pm}^{(1)}&=\omega_0 + \frac{K}{2}\pm \frac{1}{2}\sqrt{K^2+4g^2},\label{eqs4}
\end{align}
where $\theta$ is the mixing angle, defined by $\tan (2\theta)=\frac{2g}{K}$. The state $\left|n_a,n_m\right\rangle \equiv\left|n_a\right\rangle \otimes \left|n_m\right\rangle$ denotes the direct product of the Fock states with $n_a$ photons and $n_m$ magnons, respectively. It can be clearly seen in Fig.~\ref{fig:figS1} that the initially degenerate energy levels are split. This splitting is a consequence of both the interaction between the cavity and magnon modes and the Kerr nonlinearity. Indeed, the higher-energy state, $\psi_{+}^{(1)}$, corresponds to the single-excitation state of the $p_+$ polariton mode ($N_+$ = 1), while the lower-energy state, $\psi_{-}^{(1)}$, corresponds to the single-excitation state of the $p_-$ polariton mode ($N_-$ = 1). By setting $\omega_d = E_{+}^{(1)}$, we can selectively excite the higher-energy state, which corresponds to driving only the $p_+$ mode. Similarly, setting $\omega_d = E_{-}^{(1)}$ allows us to selectively drive the $p_-$ mode.

\begin{figure}[t]
    \centering
    \captionsetup[subfigure]{labelformat=empty}

    \begin{subfigure}[t]{0.49\columnwidth}
        \centering
        \includegraphics[width=\textwidth]{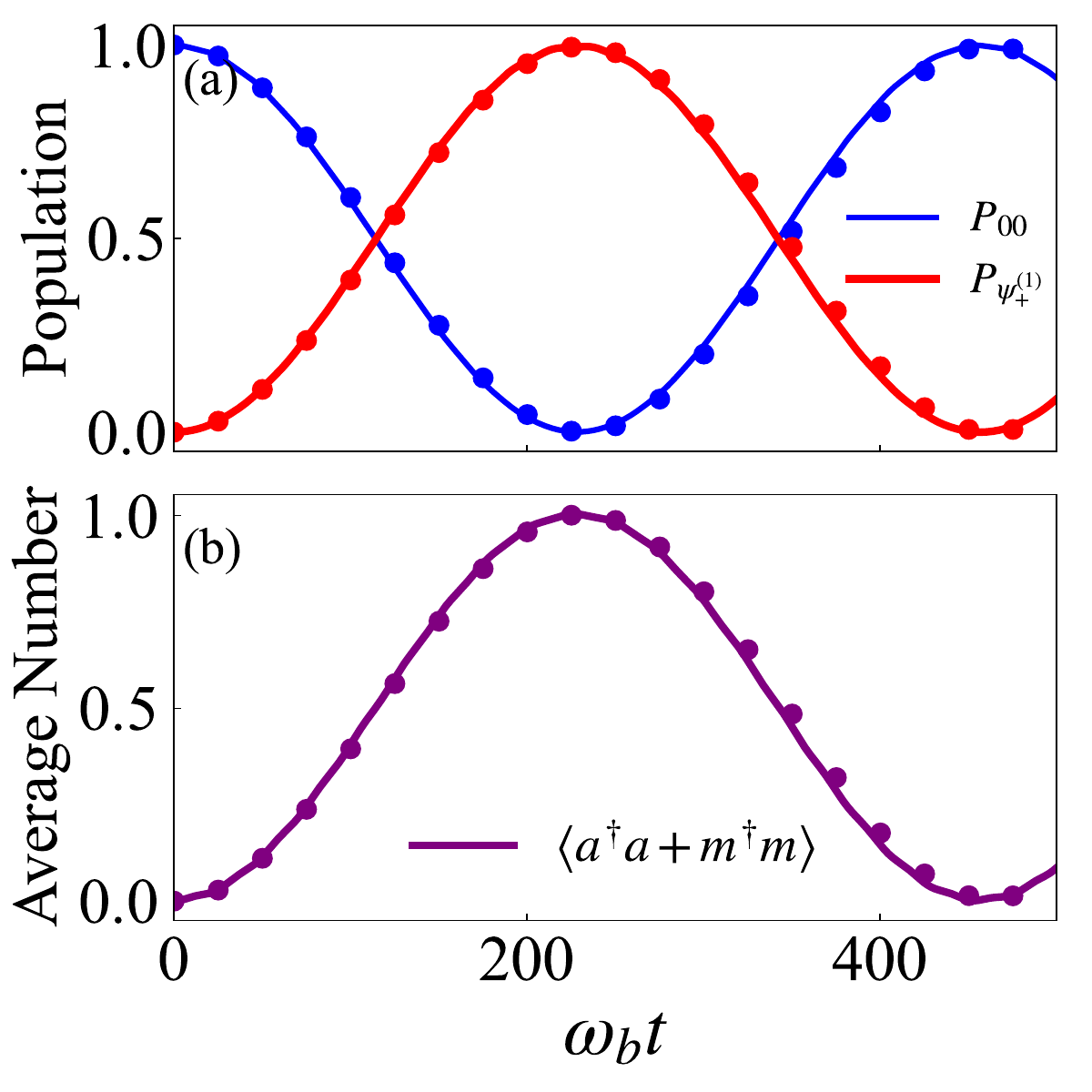}
        \caption{}
        \label{fig:figS2a}
    \end{subfigure}
    \hfill
    \begin{subfigure}[t]{0.49\columnwidth}
        \centering
        \includegraphics[width=\textwidth]{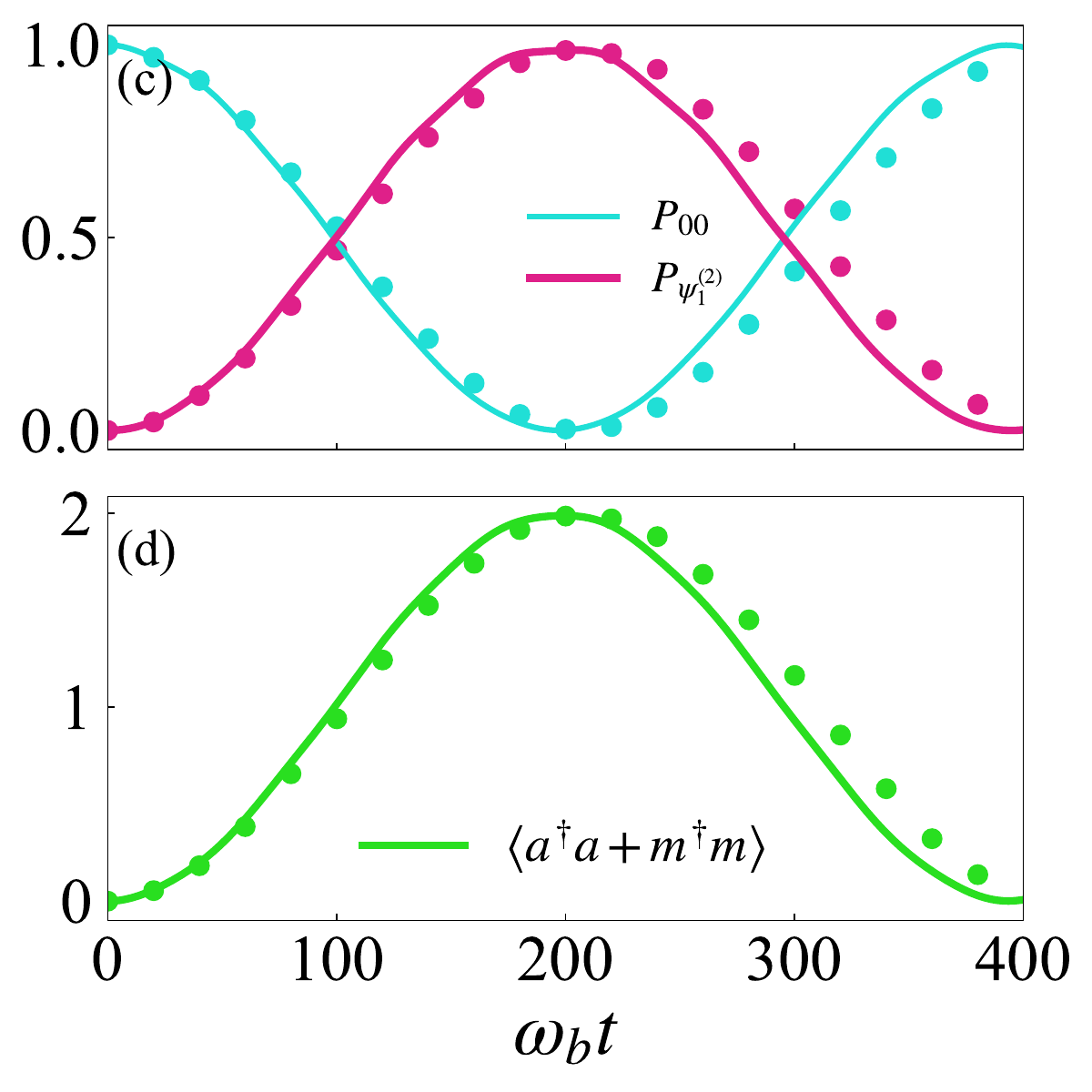}
        \caption{}
        \label{fig:figS2b}
    \end{subfigure}
    \begin{subfigure}[t]{0pt}
        \caption{}
        \label{fig:figS2c}
    \end{subfigure}

     \begin{subfigure}[t]{0pt}
        \caption{}
        \label{fig:figS2d}
    \end{subfigure}
    \vspace{-10mm}
    \caption{Time evolution of the excitation number distribution for the (a), (b) $n$ = 1 and (c), (d) $n$ = 2 cases. In (a), solid lines represent the evolution of the ground state and state $\psi_{+}^{(1)}$ governed by the original Hamiltonian (\ref{eqs1}), while dots denote the evolution of the ground state and the single-excitation $p_+$ mode state obtained from the effective Hamiltonian (\ref{eqs2}). In (b), the solid lines represent the evolution of the mean total excitation number obtained from Hamiltonian (\ref{eqs1}), while the dots denote the evolution of the mean polariton excitation number $\langle p_{+}^{\dagger} p_{+} \rangle$ calculated via the effective Hamiltonian (\ref{eqs2}). Similarly, in (c), the solid lines represent the evolution of the ground state and $\psi_{1}^{(2)}$, while the dots denote that of the ground state and the two-excitation $p_+$ mode state. System parameters are $K/g=0.05$, and $\Omega/g=G/g=0.001$. The laser frequency $\omega_d$ is chosen to satisfy the blockade condition.}
    \label{fig:figS2}
\end{figure}

For the $N$ = 2 case, we define a parameter $\lambda = E-2\omega_0$. Substituting this into the characteristic equation yields the following cubic equation for $\lambda$:
\begin{align}
\lambda^3-5K\lambda^2+(4K^2-4g^2)\lambda+8Kg^2&=0.\label{eqs5}
\end{align}
Although this equation is analytically solvable, its exact solutions are cumbersome. We therefore simplify the problem by considering the limit $g\gg K$, which allows the equation to be treated using perturbation theory. 
\begin{figure}[t]
    \centering
    \captionsetup[subfigure]{labelformat=empty}

    \begin{subfigure}[t]{0.49\columnwidth}
        \centering
        \includegraphics[width=\textwidth]{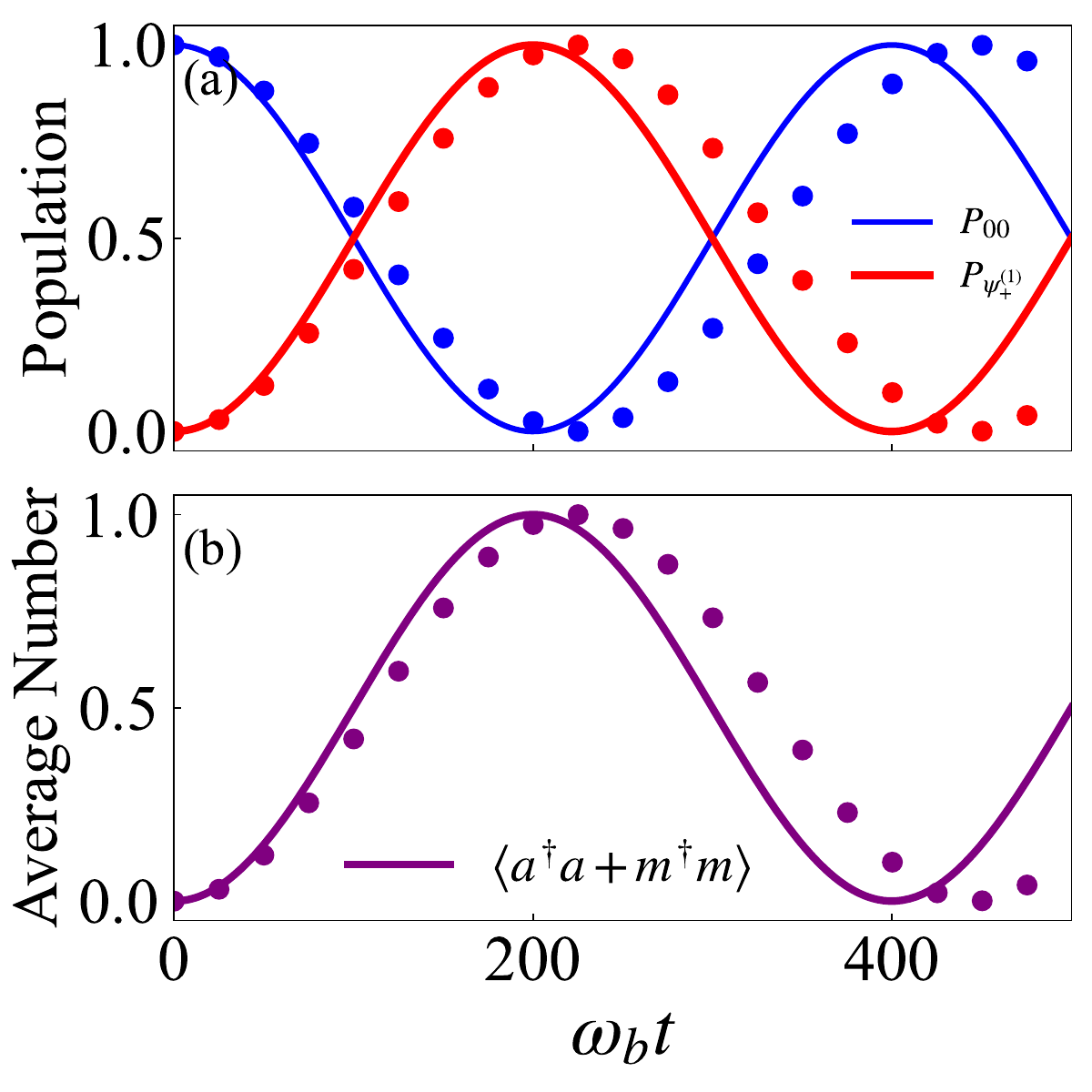}
        \caption{}
        \label{fig:figS3a}
    \end{subfigure}
    \hfill
    \begin{subfigure}[t]{0.49\columnwidth}
        \centering
        \includegraphics[width=\textwidth]{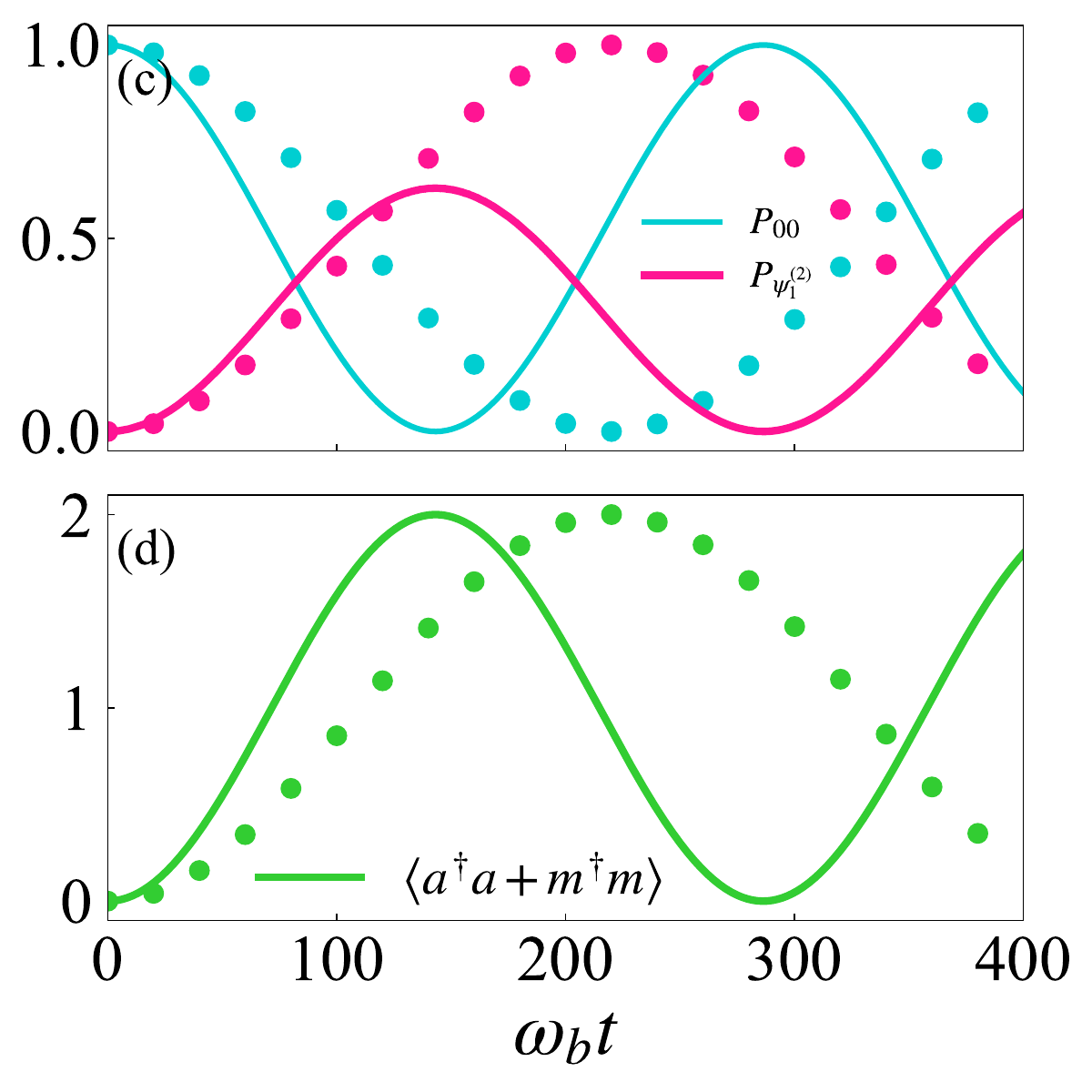}
        \caption{}
        \label{fig:figS3b}
    \end{subfigure}
    \begin{subfigure}[t]{0pt}
        \caption{}
        \label{fig:figS3c}
    \end{subfigure}

     \begin{subfigure}[t]{0pt}
        \caption{}
        \label{fig:figS3d}
    \end{subfigure}
    \vspace{-10mm}
    \caption{Time evolution of the excitation number distribution for the (a), (b) $n$ = 1 and (c), (d) $n$ = 2 cases. The solid lines and dots have the same meanings as in Fig.~\ref{fig:figS2}. System parameters are $K/g=0.5$, and $\Omega/g=G/g=0.001$. The laser frequency $\omega_d$ is chosen to satisfy the blockade condition.}
    \label{fig:figS3}
\end{figure}Assuming an ansatz for the solution of the form $\lambda = gx$ and substituting it into Eq. (\ref{eqs5}), we obtain
\begin{align}
x^3-5\epsilon x^2+(4\epsilon^2-4)x+8\epsilon&=0,\label{eqs6}
\end{align}
where $\epsilon=K/g \ll 1$. Considering the zeroth-order approximation, i.e., setting $\epsilon = 0$, we obtain the three zeroth-order solutions $x_0 = 2, 0, -2$. Next, we consider the first-order correction by setting $x = x_0 + \epsilon x_1$. Substituting this into Eq. (\ref{eqs5}) and considering the case where the zeroth-order solution is $x_0 = 2$, we solve for the first-order term and find $x_1 = 1.5$. Thus, the first approximate solution for $\lambda$ can be written as $\lambda_1 \approx 2g+1.5K$. Using the same method, the other two solutions can be found. They are $\lambda_2 \approx 2K$ and $\lambda_3 \approx -2g+1.5K$, respectively. Finally, the eigenvalues and eigenstates can be approximated as
\begin{align}
E_{1}^{(2)}& \approx 2\omega_0+2g+1.5K, \nonumber \\
E_{2}^{(2)} &\approx 2\omega_0+2K, \nonumber \\
E_{3}^{(2)}& \approx 2\omega_0-2g+1.5K, \nonumber \\
\psi_{1}^{(2)}&\approx (\frac{1}{2}+\frac{9K}{16g})\left|2,0\right\rangle+(\frac{1}{\sqrt{2}}-\frac{K}{8\sqrt{2}g})\left|1,1\right\rangle+(\frac{1}{2}-\frac{7K}{16g})\left|2,0\right\rangle,\nonumber \\
\psi_{2}^{(2)}&\approx \frac{1}{\sqrt{2}}(\left|2,0\right\rangle-\left|0,2\right\rangle)-\frac{K}{g}\left|1,1\right\rangle,\nonumber \\
\psi_{3}^{(2)}&\approx (\frac{1}{2}-\frac{9K}{16g})\left|2,0\right\rangle+(-\frac{1}{\sqrt{2}}-\frac{K}{8\sqrt{2}g})\left|1,1\right\rangle+(\frac{1}{2}+\frac{7K}{16g})\left|2,0\right\rangle.\label{eqs7}
\end{align}
It can be seen that the eigenstates consist of a $p_+$ polariton mode, a $p_-$ polariton mode, and an imperfect dark state that is slightly perturbed by the Kerr effect. By comparing Eq.~(\ref{eqs4}) and Eq.~(\ref{eqs7}), we find that for both the $p_+$ and $p_-$ polariton modes, the spacing between adjacent energy levels is non-uniform. This energy anharmonicity allows one to achieve a blockade effect by tuning the laser to be resonant with a specific transition while being off-resonant with all others. To blockade the $p_+$ mode in the $N_+ = 1$ subspace, the laser frequency must be set to be resonant with the first excited state, i.e., $\omega_d = E_{+}^{(1)}$. In the limit where $g\gg K$, the blockade condition is found to be $\omega_d =\omega_0 +g + \frac{K}{2}$. Similarly, to achieve a blockade in the $N_+ = 2$ subspace, the drive frequency must satisfy the condition $\omega_d = \frac{E_1^{(2)}}{2}=\omega_0+g+\frac{3K}{4}$. We observe that the blockade condition derived directly from the original Hamiltonian is in perfect agreement with the one derived in the main text from the effective Hamiltonian. This confirms that the original Hamiltonian and the effective model we presented in the main text are fully equivalent for describing this phenomenon. To numerically illustrate this, we compare the system evolution under both the original and effective Hamiltonians in Fig.~\ref{fig:figS2}. As shown in the figures, the two Hamiltonians yield nearly identical results in the regime $g\gg K$. As $K$ increases and the condition $g\gg K$ is no longer satisfied, the validity of the effective Hamiltonian gradually deteriorates, as illustrated in Fig.~\ref{fig:figS3}. However, a comparison of Figs.~\ref{fig:figS2} and~\ref{fig:figS3} shows that the maximum total excitation number of the system does not change with increasing $K$; the system remains in the blockade regime. This is because increasing $K$ does not alter the blockade condition, but rather increases the effective Rabi frequency, thereby shortening the oscillation period between the ground state and the eigenstates. Meanwhile, for the $n=2$ case, the increase of $K$ renders the first‑order perturbation to the eigenwave function in Eq.~(\ref{eqs7}) no longer valid, leading to a deviation of $\psi_{1}^{(2)}$ in Fig.~\ref{fig:figS3}\subref{fig:figS3c} from the true system wave function.

\begin{figure}[t]
    \centering
    \captionsetup[subfigure]{labelformat=empty}

    \begin{subfigure}[t]{0.49\columnwidth}
        \centering
        \includegraphics[width=\textwidth]{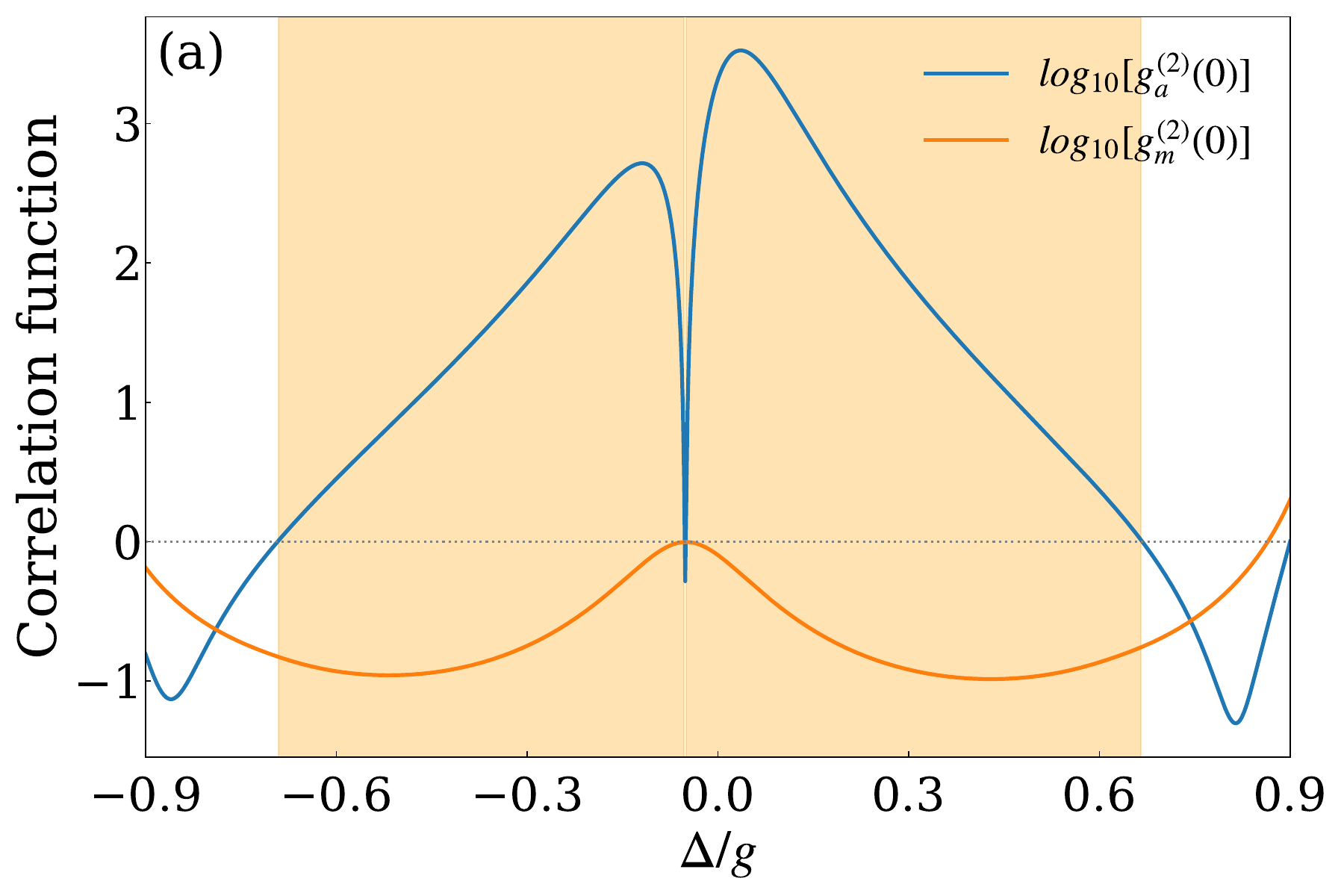}
        \caption{}
        \label{fig:figS4a}
    \end{subfigure}
    \hfill
    \begin{subfigure}[t]{0.49\columnwidth}
        \centering
        \includegraphics[width=\textwidth]{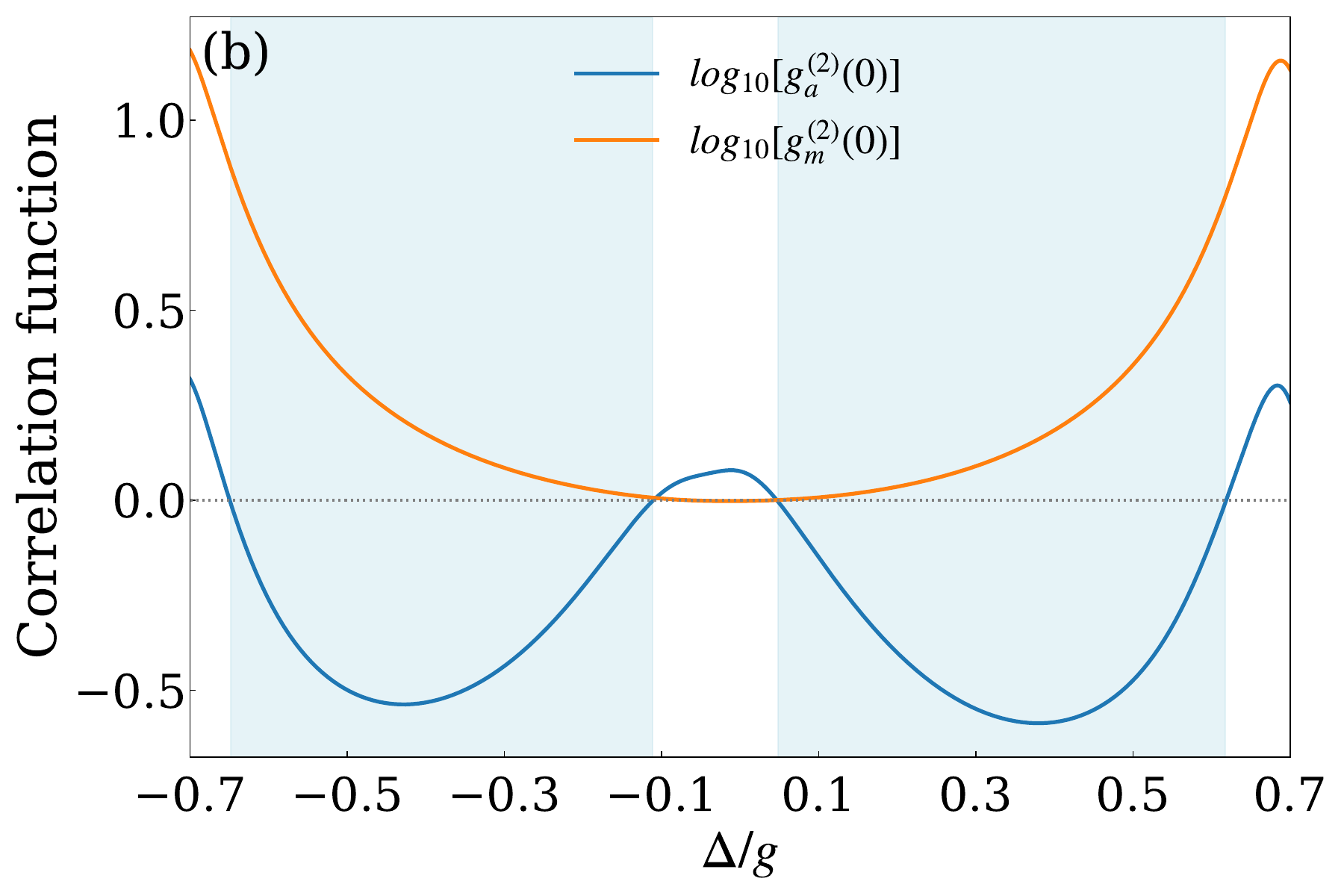}
        \caption{}
        \label{fig:figS4b}
    \end{subfigure}
\begin{subfigure}[t]{0.49\columnwidth}
        \centering
        \includegraphics[width=\textwidth]{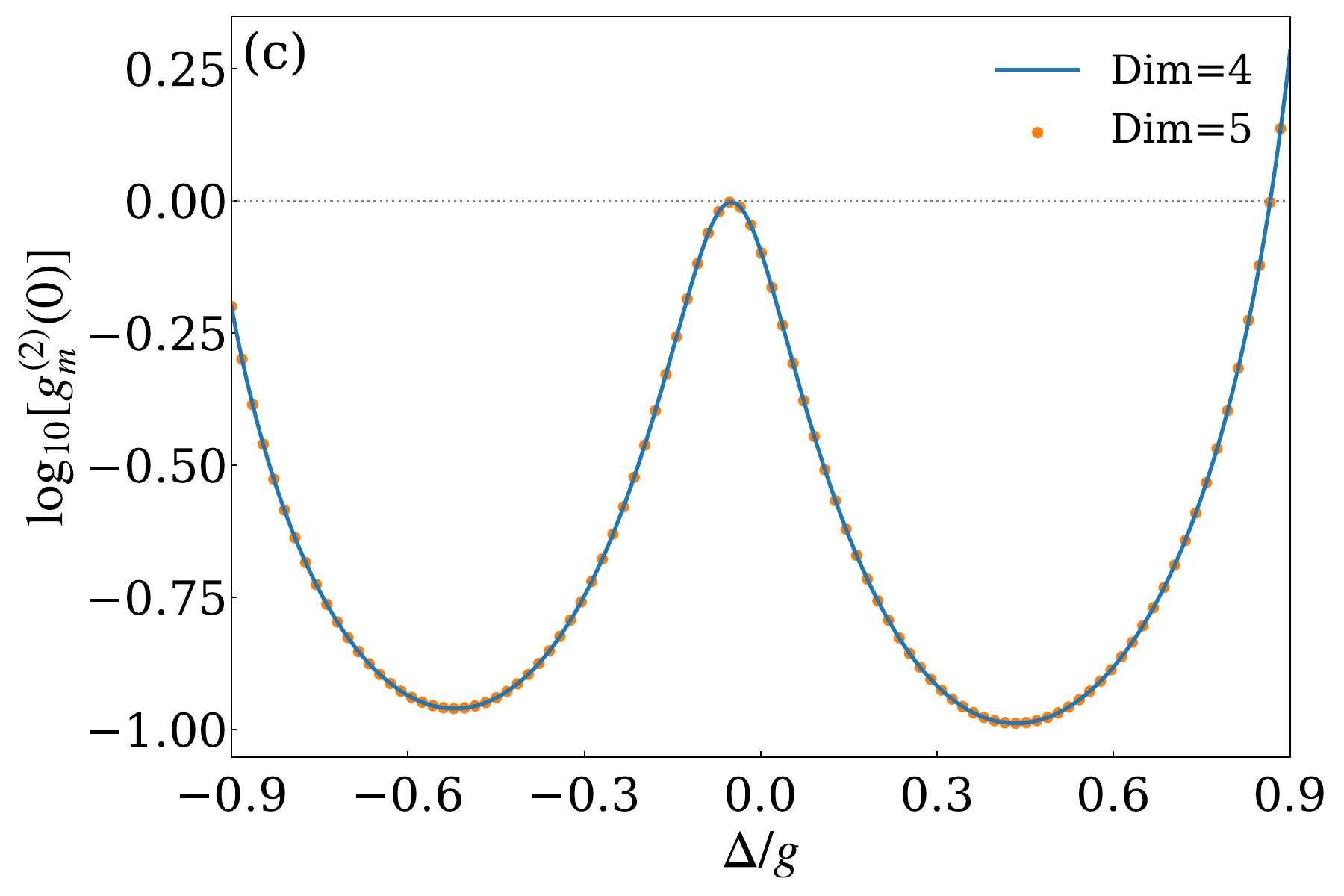}
        \caption{}
        \label{fig:figS4c}
    \end{subfigure}
    \hfill
    \begin{subfigure}[t]{0.49\columnwidth}
        \centering
        \includegraphics[width=\textwidth]{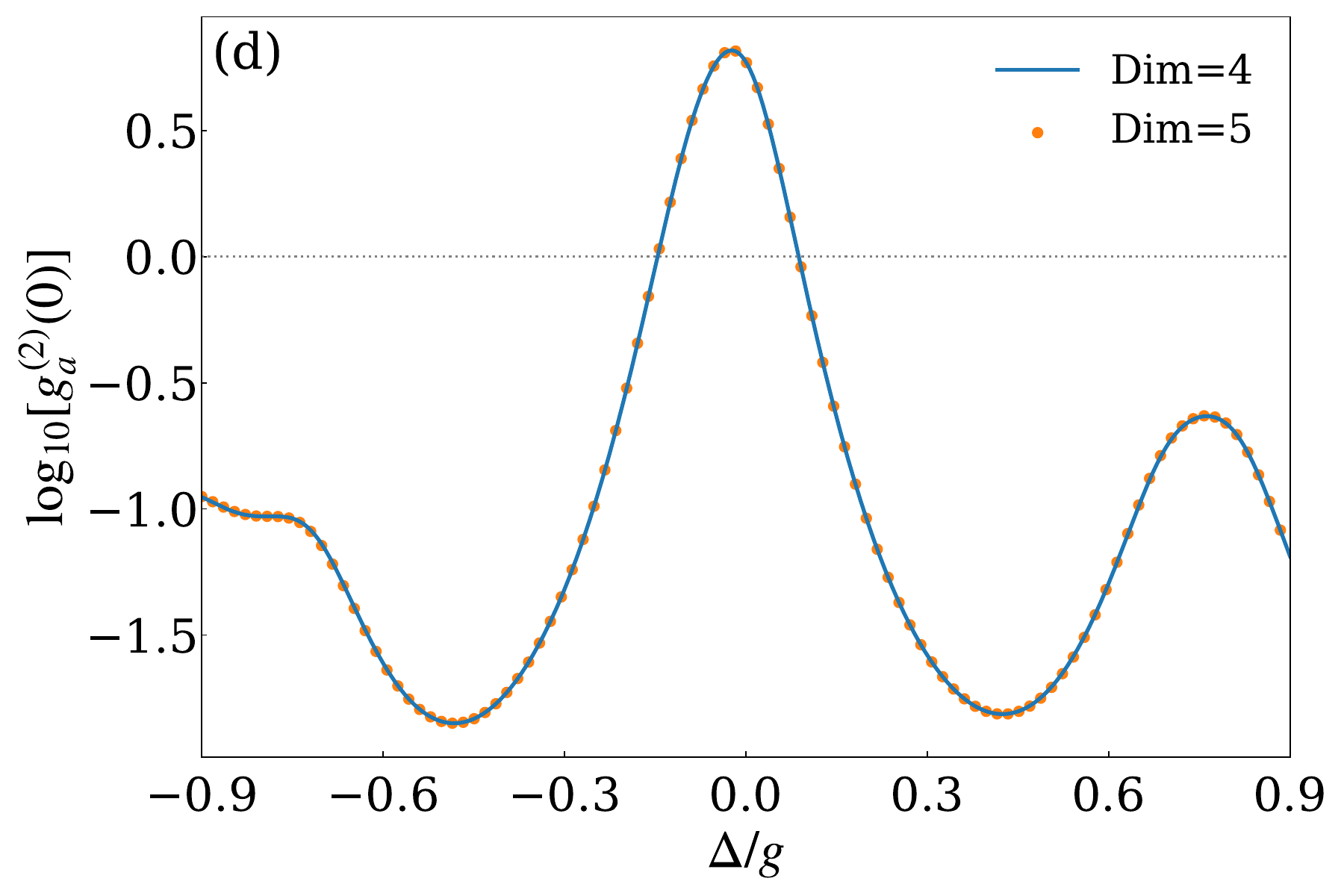}
        \caption{}
        \label{fig:figS4d}
    \end{subfigure}
    \vspace{-8mm}
    \caption{Second-order magnon and photon correlation functions on a logarithmic scale $\log_{10}[ g_m^{(2)}(0)]$ and $\log_{10}[ g_a^{(2)}(0)]$ as a	function of $\Delta/g$. The orange region in panel (a) denotes the UMB regime, while the light-blue region in panel (b) denotes the UPB regime. Blue lines and orange dots in panels (c) and (d) indicate that the Hilbert space dimension of each mode is truncated to $\text{Dim}=4$ and 5, respectively. System parameters are $K/g=0.05$, $\Omega/g=0.01$, $\kappa_{a}/g=\kappa_{m}/g=\kappa/g=0.2$. The values of $G$ in panels (a) and (c) satisfy Eq.~(\ref{eqs26}), whereas those in panels (b) and (d) satisfy Eq.~(\ref{eqs28}).}
    \label{fig:figS4}
\end{figure}

\hypertarget{app:B}{}
\section{Single-photon and Single-magnon Blockades}
The blockade effect in the dissipative system can also be characterized by the equal-time correlation function. The photon blockade (PB) and magnon blockade (MB) can be quantified by the equal-time second order correlation function
\begin{align}
g^{(2)}_{a}(0) &= \frac{\langle \hat{a}^{\dagger 2} \hat{a}^2 \rangle} {\langle \hat{a}^{\dagger} \hat{a} \rangle^2 }
=\frac{\mathrm{Tr}(\rho_{ss}\hat{a}^{\dagger 2} \hat{a}^2) } {[\mathrm{Tr}(\rho_{ss}\hat{a}^{\dagger} \hat{a})]^2 } ,\\
g^{(2)}_{m}(0) &= \frac{\langle \hat{m}^{\dagger 2} \hat{m}^2 \rangle} {\langle \hat{m}^{\dagger} \hat{m} \rangle^2}
=\frac{\mathrm{Tr}(\rho_{ss}\hat{m}^{\dagger 2} \hat{m}^2) } {[\mathrm{Tr}(\rho_{ss}\hat{m}^{\dagger} \hat{m})]^2 }. \label{eqs8}
\end{align}
The criterion for the PB and MB is that the corresponding correlation functions satisfy \cite{PhysRevLett.118.133604,PhysRevLett.121.153601,PhysRevA.102.053710,Zhang:25}
\begin{align}
g^{(2)}_{a}(0)  <1, g^{(2)}_{m}(0)  <1,\label{eqs9}
\end{align}
which indicates that both photons and magnons are antibunched. Next, we investigate the unconventional photon and magnon blockade. The Hamiltonian of the proposed system is written as 
\begin{align}
H = \Delta_{c}a^{\dagger}a+\Delta _{m}m^{\dagger}m + K(a^{\dagger}a)^{2} + g(a^{\dagger}m+am^{\dagger})+\Omega(a^{\dag} +a)+G(a^{\dag 2} +a^{2}).
\end{align}
For simplicity, we set $\Delta_c = \Delta_m = \Delta$ in the following. To derive the optimal parameter conditions, we start from the Schr\"{o}dinger equation, $i \partial |\psi\rangle/\partial t = H_{\text{non}} |\psi\rangle$ , where
\begin{align}
H_{\text{non}}  = H - i\frac{\kappa_a}{2}a^{\dagger}a -i\frac{\kappa_m}{2}m^{\dagger}m \label{eqs10}
\end{align}
is the non-Hermitian Hamiltonian of the system and $|\psi\rangle$ represents the wave function. In the weak-driving regime ($\Omega,G \ll \kappa_a,\kappa_m$), the Hilbert space of the system can be truncated to the two-excitation subspace. Consequently, the steady-state wave function is given by
\begin{align}
|\psi\rangle = C_{00}|00\rangle + C_{10}|10\rangle + C_{01}|01\rangle+ C_{20}|20\rangle + C_{11}|11\rangle + C_{02}|02\rangle.\label{eqs11}
\end{align}
Here, $|am\rangle = |a\rangle \otimes |m\rangle$ is the tensor product state of the cavity mode $a$ and the magnon mode $m$, with $C_{am}$ being the corresponding probability amplitude. Substituting Eqs. (\ref{eqs10}) and (\ref{eqs11}) into the Schr\"{o}dinger equation, we obtain the dynamical equations for the probability amplitudes
\begin{align}
i \dot{C}_{10} &= (\tilde{\Delta}_a+K) C_{10} + g C_{01} + \Omega C_{00} + \sqrt{2} \Omega C_{20}, \\
i \dot{C}_{01} &= \tilde{\Delta}_m  C_{01} + g C_{10} + \Omega C_{11}, \\
i \dot{C}_{20} &= (2 \tilde{\Delta}_a+4K) C_{20} + \sqrt{2} g C_{11} + \sqrt{2} \Omega C_{10} + \sqrt{2} G C_{00}, \\
i \dot{C}_{11} &= (\tilde{\Delta}_a + \tilde{\Delta}_m + K) C_{11} + \sqrt{2} g (C_{20} + C_{02}) + \Omega C_{01}, \\
i \dot{C}_{02} &= 2 \tilde{\Delta}_m C_{02} + \sqrt{2} g C_{11},
\end{align}
where $\tilde{\Delta}_a=\Delta-i\kappa_a/2$ and $\tilde{\Delta}_m=\Delta-i\kappa_m/2$. In the weak-driving limit, $C_{00} \approx 1 \gg C_{10}, C_{01} \gg C_{11}, C_{02}, C_{20}$, allowing us to ignore its dynamics. By further disregarding $C_{11}$ and $C_{20}$ in the first two equations, we obtain
\begin{align}
C_{01} &= -\frac{g \Omega}{g^{2} - (\tilde{\Delta}_{a}+K) \tilde{\Delta}_{m}}, \\
C_{10} &= \frac{\Omega \tilde{\Delta}_m}{g^{2} - (\tilde{\Delta}_{a}+K) \tilde{\Delta}_{m} }.
\end{align}
\begin{figure}[t]
    \centering
    \captionsetup[subfigure]{labelformat=empty}
    \begin{subfigure}[t]{0.6\columnwidth}
        \centering
        \includegraphics[width=\textwidth]{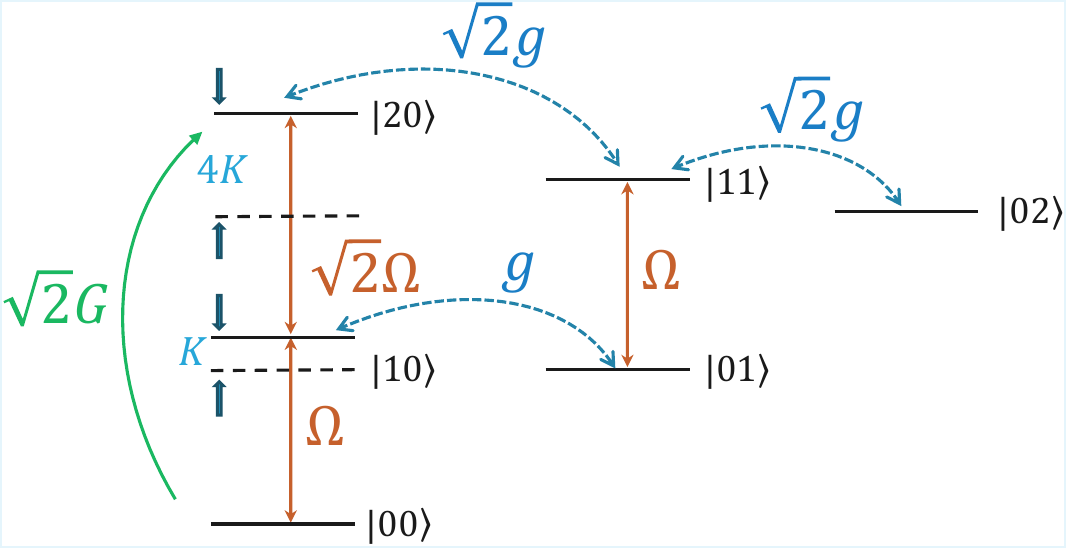}
        \caption{}

    \end{subfigure}

    \caption{Energy-level transition diagram of the system in the two-excitation subspace. Destructive interference between different transition pathways suppresses the population of the two-photon and two-magnon states.}
    \label{fig:figS5}
\end{figure}
Substituting $C_{01}$ and $C_{10}$ into the remaining three equations yields
\begin{align}
C_{20} &= \frac{GD(\tilde{\Delta}_m^2 - D) + \Omega^2 \tilde{\Delta}_m^2(\tilde{\Delta}_{a}+ \tilde{\Delta}_{m} + K) }{\sqrt{2}D \left[ (\tilde{\Delta}_a + \tilde{\Delta}_m + 2K)D - K \tilde{\Delta}_m^2 \right]},\\
C_{02} &= \frac{g^{2} \left[ GD 
            + (\tilde{\Delta}_{a} + 2K + \tilde{\Delta}_{m} ) \Omega^{2} \right]}
            {\sqrt{2}D \left[ (\tilde{\Delta}_{a} + \tilde{\Delta}_{m} + 2K)
            D -K\tilde{\Delta}_{m}^2 \right]},
\end{align}
where $D=g^{2} -  \tilde{\Delta}_{m}(\tilde{\Delta}_{a}+ K)$. Thus, we obtain the analytical expressions for the two equal-time second-order correlation functions
\begin{align}
g_a^{(2)}(0)&= \frac{2|C_{20}|^2}{\bigl(|C_{10}|^2 + 2|C_{20}|^2 + |C_{11}|^2\bigr)^2} 
               \approx \frac{2|C_{20}|^2}{|C_{10}|^4},\\
g_m^{(2)}(0)&= \frac{2|C_{02}|^2}{\bigl(|C_{01}|^2 + 2|C_{02}|^2 + |C_{11}|^2\bigr)^2} 
               \approx \frac{2|C_{02}|^2}{|C_{01}|^4}. 
\end{align}
To achieve $g_a \rightarrow 0$, the condition $C_{20}=0$ must be satisfied. In this case, the photons exhibit a pronounced antibunching effect, implying that a near-perfect PB can be achieved. Likewise, achieving MB requires $C_{02}=0$. First, we consider the case of MB. By setting $C_{02}=0$, the equation can be decomposed into its real and imaginary parts:
\begin{align}
G\left[ g^{2} - \Delta^2  + \frac{\kappa_{a} \kappa_{m}}{4}-\Delta K \right] 
+ 2\Omega^{2}(\Delta + K) &= 0,\label{eqs26}\\ 
G(\frac{\kappa_{a} +\kappa_{m}}{2}+\frac{\kappa_{m}}{2}K)
- \frac{\Omega^2}{2}(\kappa_{a}+\kappa_{m}) &= 0. 
\end{align}
In the limits of strong coupling and weak driving, the imaginary part is much smaller than the real part. Thus, it suffices to consider only the real-part equation. At this point, we have successfully derived the optimal parameter conditions for unconventional magnon blockade (UMB), as expressed in Eq.~(\ref{eqs26}). For unconventional photon blockade (UPB), the imaginary parts of the equations are non-negligible. We therefore directly present the conditions for UPB as follows:
\begin{align}
GD(\tilde{\Delta}_m^2 - D) + \Omega^2 \tilde{\Delta}_m^2(\tilde{\Delta}_{a}+ \tilde{\Delta}_{m} + K)= 0. \label{eqs28}
\end{align}
Note that, in principle, $G$ in Eq.~(\ref{eqs28}) is complex and can be written as $G=|G|e^{i\varphi}$, where $\varphi$ is the phase of the pump laser. 
However, we assume that, regardless of the system parameters, $\varphi$ always takes the value that satisfies Eq.~(\ref{eqs28}). 
Therefore, $G$ is treated as a real parameter without distinguishing between $G$ and $|G|$, which facilitates a clearer comparison between the UMB and UPB cases. 
This treatment is adopted throughout the subsequent analysis and numerical simulations.

In Figs.~\ref{fig:figS4}\subref{fig:figS4a} and~\ref{fig:figS4}\subref{fig:figS4b}, we plot the second-order correlation functions as a function of $\Delta/g$ under the optimal parameter conditions for UMB and UPB, respectively. It can be observed that the UMB region is mainly concentrated within the interval $\Delta/g\in[-0.69,0.67]$, whereas the UPB region is distributed among four separated areas. With $G$ set to its optimal value, the optimal parameter for UMB occurs at $\Delta/g=0.43$, while that for UPB appears at $\Delta/g=-0.48$. Therefore, by appropriately tuning $\Delta$ and $G$, one can effectively realize the desired single-photon blockade or single-magnon blockade.

In the preceding discussion, we restricted the system to the two-excitation subspace; accordingly, in the numerical simulations, the Hilbert space dimension of each mode is truncated to $\text{Dim}=4$. To verify the validity of the numerical simulations, Figs.~\ref{fig:figS4}\subref{fig:figS4c} and~\ref{fig:figS4}\subref{fig:figS4d} plot $\log_{10}[ g_m^{(2)}(0)]$ and $\log_{10}[ g_a^{(2)}(0)]$ for different truncation dimensions $\text{Dim}$. It can be seen that the correlation functions for $\text{Dim}=4$ and 5 are almost identical, indicating that increasing the Hilbert space dimension of each mode has no impact on the steady-state correlation functions of the system.

In the main text, we state that the unconventional blockade originates from destructive interference between different transition pathways. Specifically, as shown in Fig.~\ref{fig:figS5}, for UPB there are three transition pathways that can lead to the two-photon state: (i) $|00\rangle \rightarrow |10\rangle \xrightarrow{\sqrt{2}\Omega} |20\rangle$, (ii) $|00\rangle \xrightarrow{\sqrt{2}G} |20\rangle$, and (iii) $|00\rangle \rightarrow |10\rangle \rightarrow |01\rangle \rightarrow |11\rangle \xrightarrow{\sqrt{2}g} |20\rangle$. When the system parameters satisfy Eq.~(\ref{eqs28}), these three transition pathways undergo destructive interference, resulting in a nearly vanishing population of the two-photon state.  Similarly, for UMB, there are three transition pathways that can lead to the two-magnon state: (i) $|00\rangle \rightarrow |10\rangle \rightarrow |01\rangle \rightarrow |11\rangle \xrightarrow{\sqrt{2}g} |02\rangle$, (ii) $|00\rangle \rightarrow |10\rangle \xrightarrow{\sqrt{2}\Omega} |20\rangle \rightarrow |11\rangle \rightarrow |02\rangle$, and (iii) $|00\rangle \xrightarrow{\sqrt{2}G} |20\rangle \rightarrow |11\rangle \xrightarrow{\sqrt{2}g} |02\rangle$. When the system parameters satisfy Eq.~(\ref{eqs26}), these three transition pathways undergo destructive interference, leading to a nearly vanishing population of the two-magnon state.

\begin{figure}[t]
    \centering
    \captionsetup[subfigure]{labelformat=empty}

    \begin{subfigure}[t]{0.4\columnwidth}
        \centering
        \includegraphics[width=\textwidth]{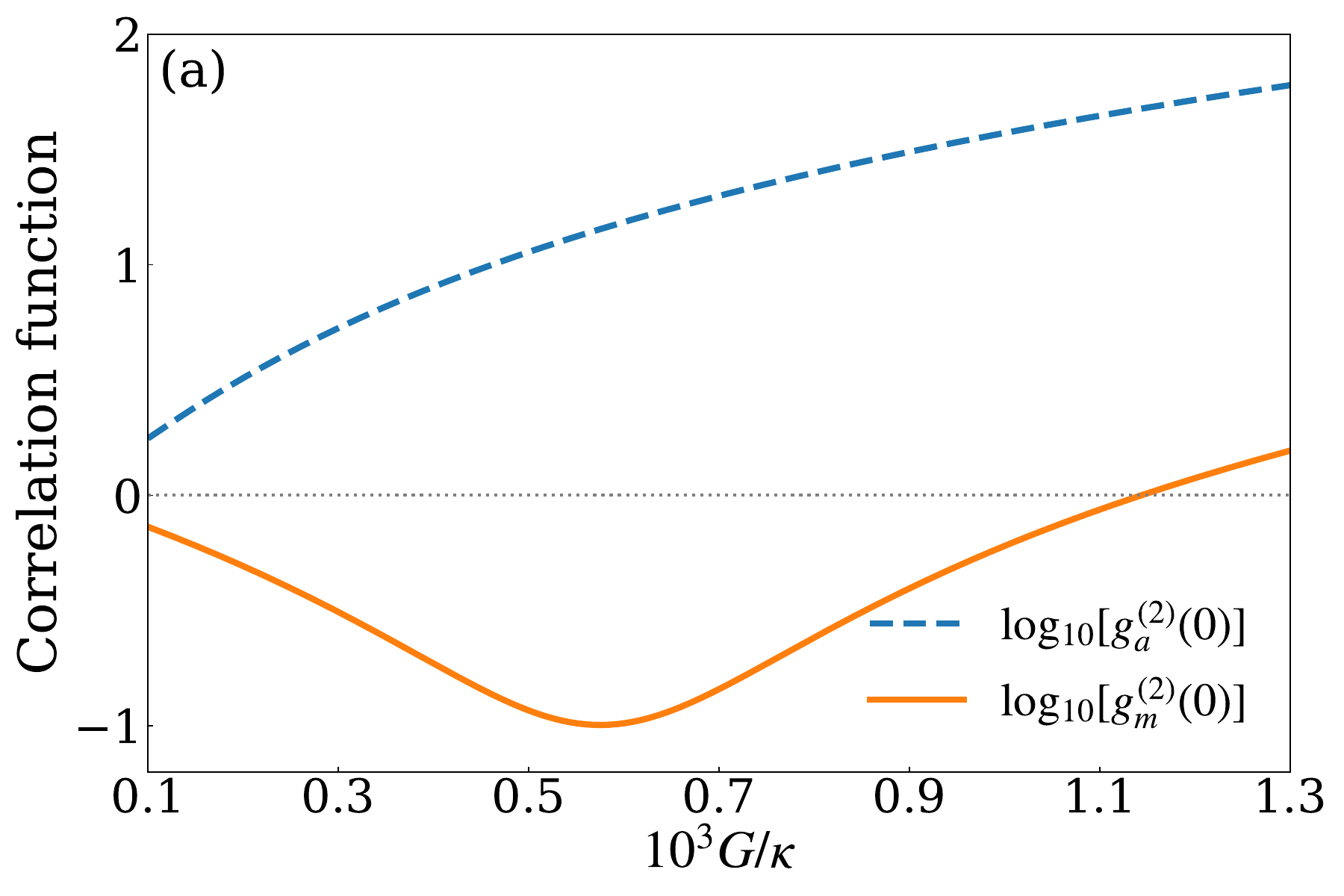}
        \caption{}
        \label{fig:figS6a}
    \end{subfigure}
    \hfill
    \begin{subfigure}[t]{0.4\columnwidth}
        \centering
        \includegraphics[width=\textwidth]{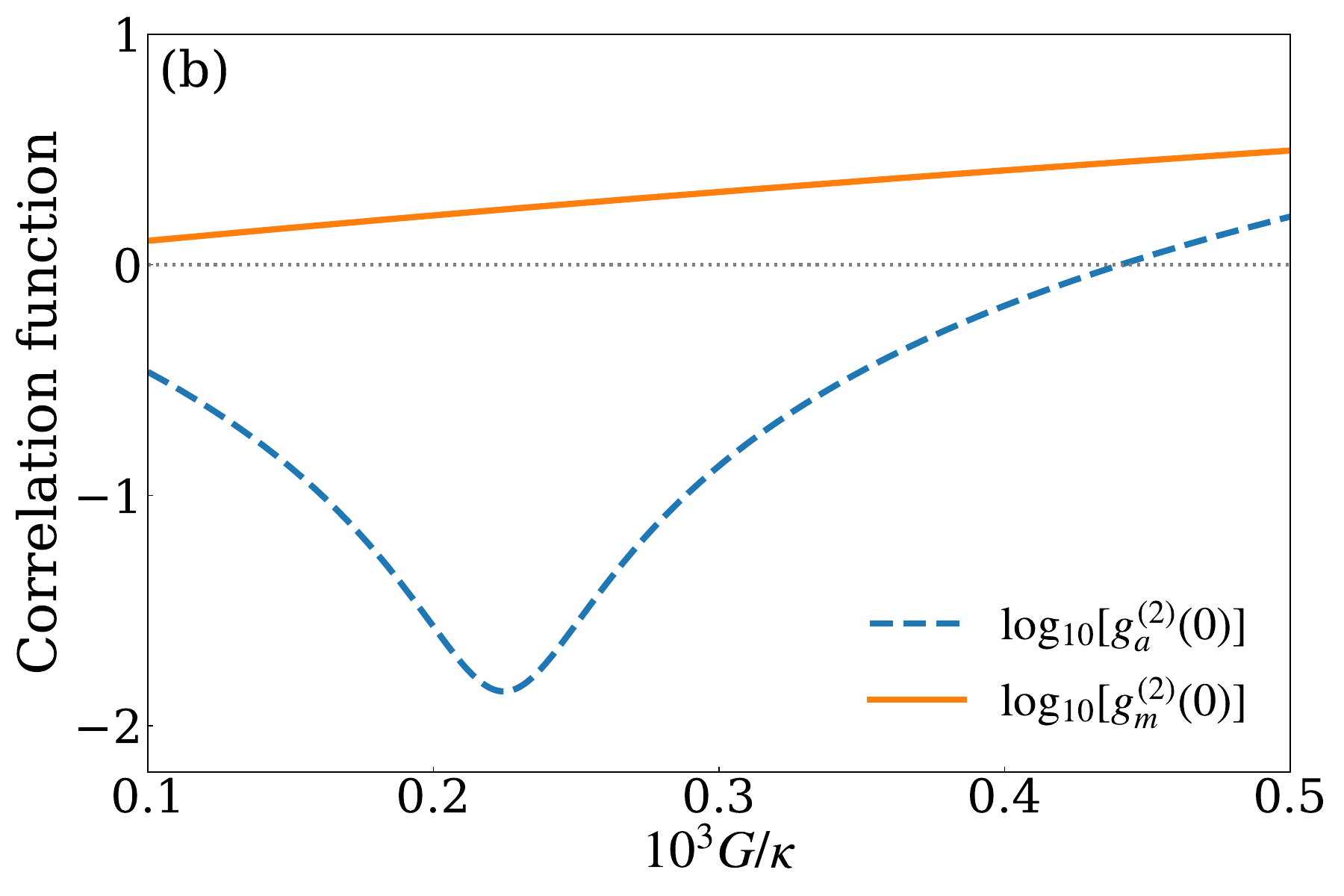}
        \caption{}
        \label{fig:figS6b}
    \end{subfigure}
\begin{subfigure}[t]{0.4\columnwidth}
        \centering
        \includegraphics[width=\textwidth]{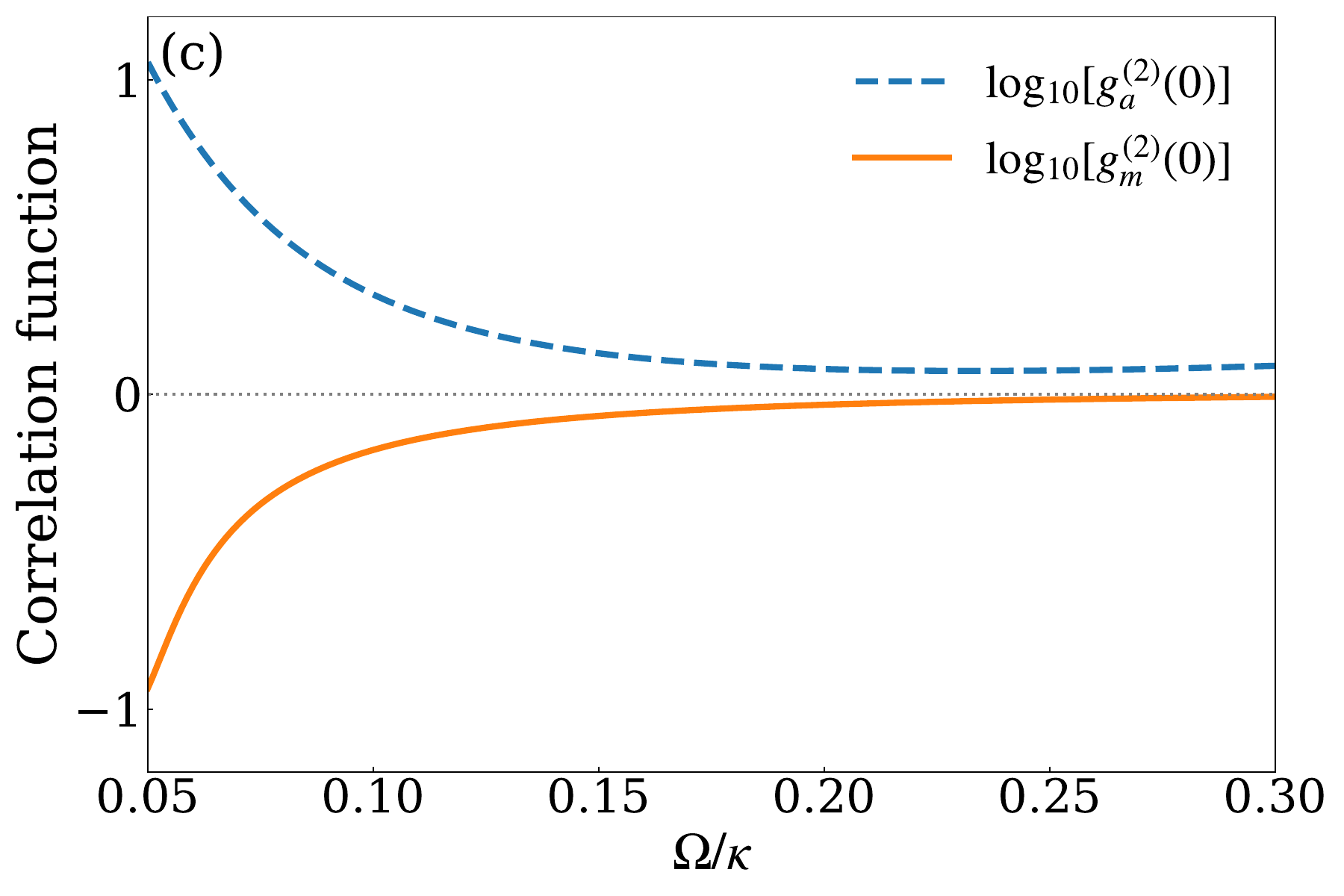}
        \caption{}
        \label{fig:figS6c}
    \end{subfigure}
    \hfill
    \begin{subfigure}[t]{0.4\columnwidth}
        \centering
        \includegraphics[width=\textwidth]{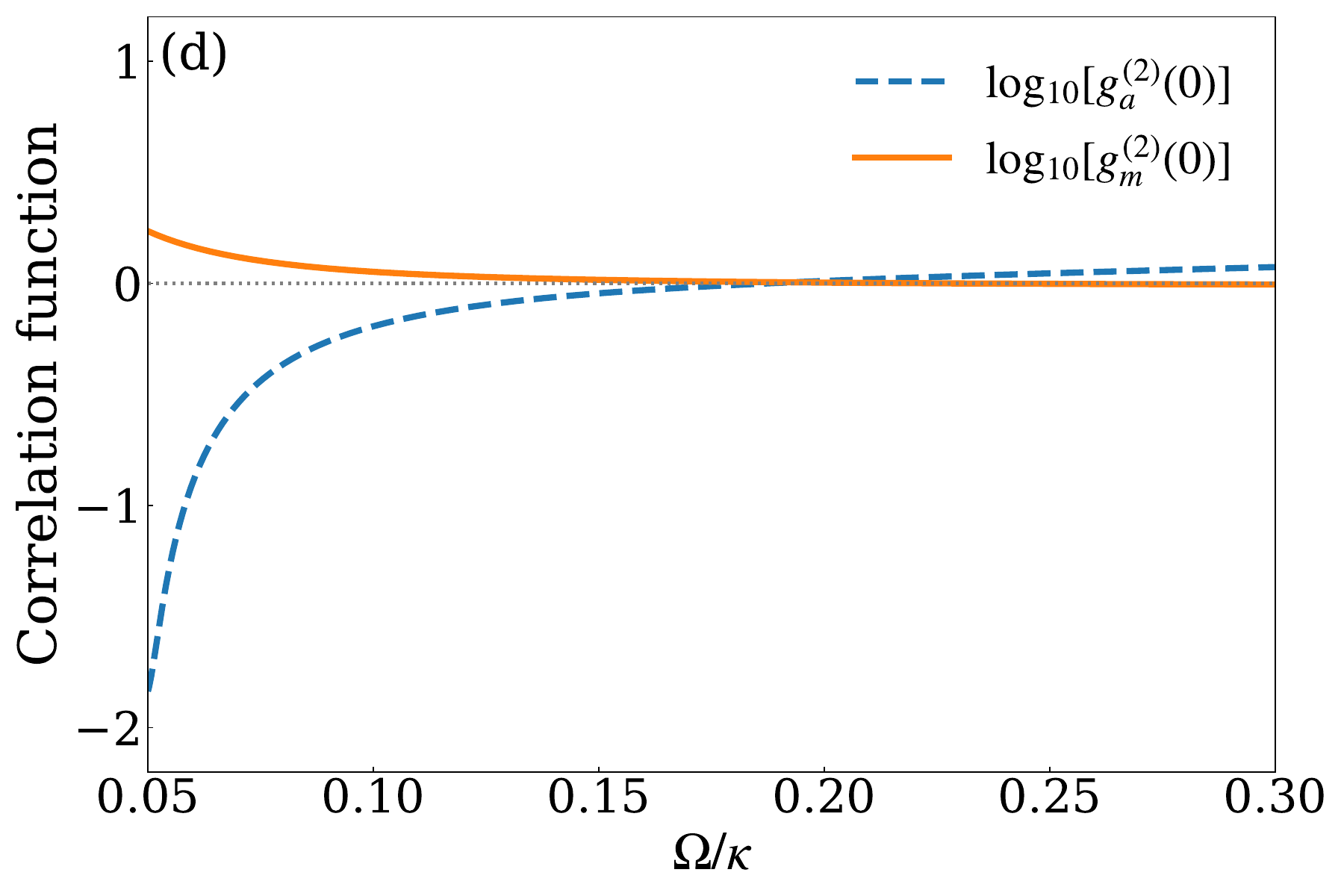}
        \caption{}
        \label{fig:figS6d}
    \end{subfigure}
    \vspace{-8mm}
    \caption{$\log_{10}[g_{m}^{(2)}(0)]$ and $\log_{10}[g_{a}^{(2)}(0)]$ as a function of $G/\kappa$ in panels (a) and (b), and $\Omega/\kappa$ in panels (c) and (d). Panels (a) and (c) correspond to the UMB case, while panels (b) and (d) correspond to the UPB case. $K/g=0.05$ and $\kappa_{a}/g=\kappa_{m}/g=\kappa/g=0.2$. $\Omega/\kappa=0.05$ in panels (a) and (b), $G/\kappa=5\times 10^{-4}$ in panel (c) and $G/\kappa=2.2\times 10^{-4}$ in panel (d). $\Delta/g=0.43$ in panels (a) and (c), and $-0.48$ in panels (b) and (d).}
    \label{fig:figS6}
\end{figure}

Finally, we discuss the effects of the two driving strengths, $\Omega$ and $G$, on UPB and UMB. As shown in Fig.~\ref{fig:figS6}, the unconventional blockade effect eventually disappears as $G$ and $\Omega$ increase, because the weak-driving condition discussed earlier is no longer satisfied and the system is excited to states with a total particle number exceeding 1. However, in Figs.~\ref{fig:figS6}\subref{fig:figS6a} and~\ref{fig:figS6}\subref{fig:figS6b}, the correlation functions corresponding to the blocked modes initially decrease as $G$ increases, reach a minimum, and then increase again with further increase of $G$. This can be explained as follows: under the parameters we use, when $G=0$, the conditions for unconventional blockade---i.e., Eqs.~(\ref{eqs26}) and~(\ref{eqs28})--- are not satisfied. In this case, the system is driven solely by $\Omega$ and thus exhibits the characteristics of a coherent light field. As $G$ increases from zero, the system begins to satisfy the conditions for unconventional blockade, causing the corresponding correlation functions to decrease.

\hypertarget{app:C}{}
\section{Optimal Drive Strength Parameter Settings}
\begin{figure}[t]
    \centering
    \captionsetup[subfigure]{labelformat=empty}

    \begin{subfigure}[t]{0.8\columnwidth}
        \centering
        \includegraphics[width=\textwidth]{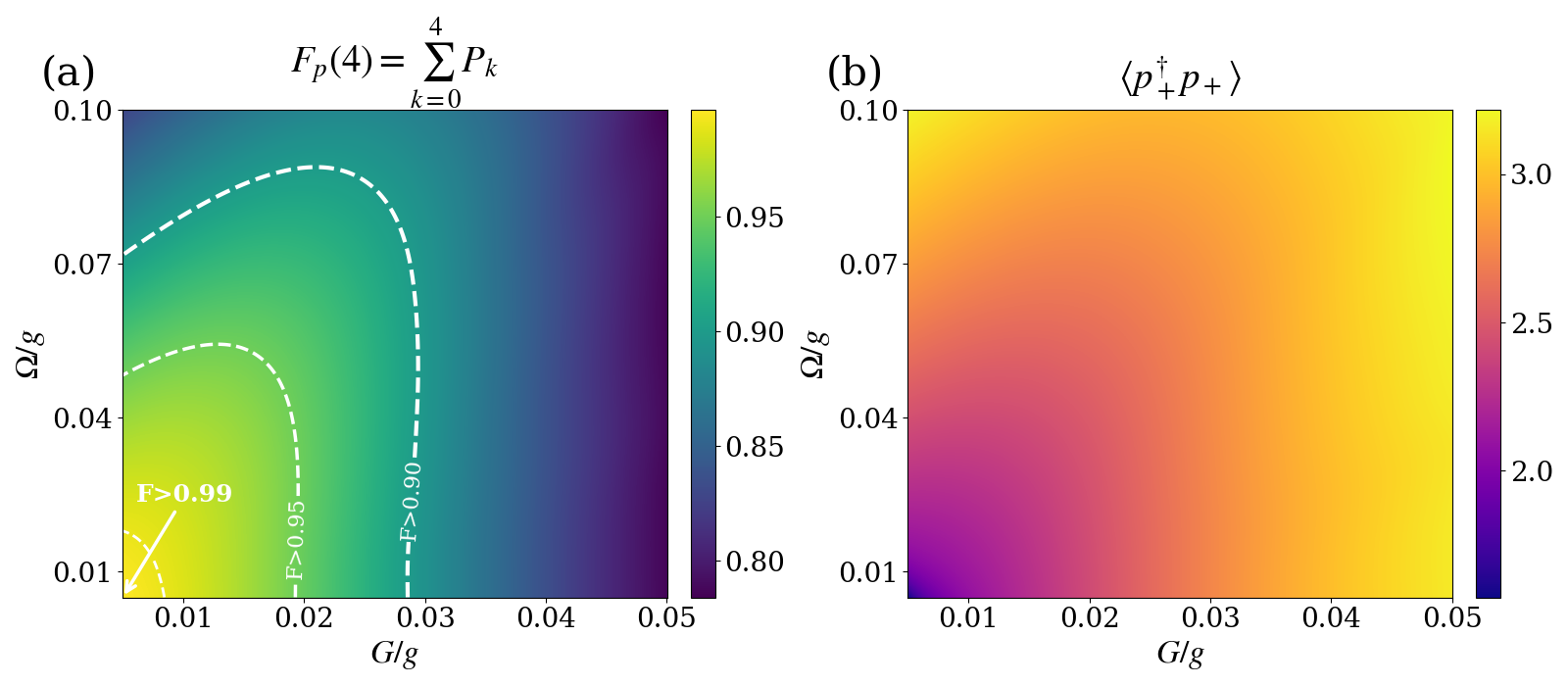}
        \caption{}
        \label{fig:figS7a}
    \end{subfigure}
    \hfill
    \begin{subfigure}[t]{0.8\columnwidth}
        \centering
        \includegraphics[width=\textwidth]{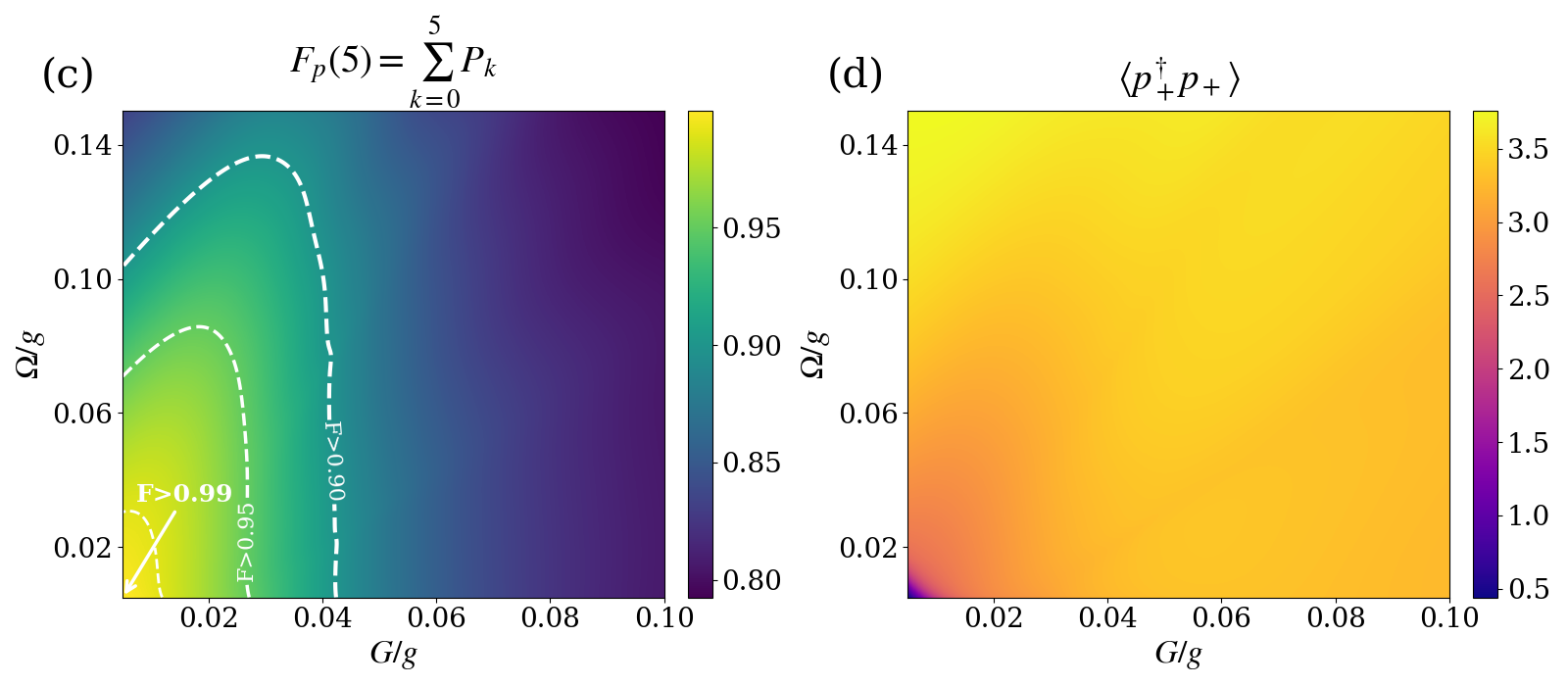}
        \caption{}
        \label{fig:figS7b}
    \end{subfigure}
    \begin{subfigure}[t]{0pt}
        \caption{}
        \label{fig:figS7c}
    \end{subfigure}

     \begin{subfigure}[t]{0pt}
        \caption{}
        \label{fig:figS7d}
    \end{subfigure}
    \vspace{-8mm}
    \caption{(a) Fidelity $F_p(4)$ and (b) average polariton excitation number $\langle p_{+}^{\dagger} p_{+} \rangle$ as functions of $G/g$ and $\Omega/g$ under 4CMPB. (c) Fidelity $F_p(5)$ and (d) average polariton excitation number $\langle p_{+}^{\dagger} p_{+} \rangle$ as functions of $G/g$ and $\Omega/g$ under 5CMPB. The white dashed lines in panels (a) and (c) mark the regions where the fidelity $F_p>0.9$, $F_p>0.95$, and $F_p>0.99$, respectively. System parameters are $K/g=0.05$ and $\kappa/g=0.001$.}
    \label{fig:figS7}
\end{figure}

In Fig.~\ref{fig:fig5} of the main text, we have discussed, for CMPB of different orders in the presence of dissipation, the dependence of the fidelity and the average polariton excitation number on the drive strength. We conclude that dissipation can, to some extent, assist the blockade effect, enabling high‑quality CMP blockade to be maintained even under relatively strong driving. Here, taking 4CMPB and 5CMPB as examples, we additionally discuss the choice of the optimal drive strength. As shown in Figs.~\ref{fig:figS7}\subref{fig:figS7a} and~\ref{fig:figS7}\subref{fig:figS7c}, the fidelity is more sensitive to the two-photon drive strength. This is readily understood, as the two‑photon drive more efficiently pumps the system into higher excited states. Moreover, higher-order blockade can tolerate larger driving strengths while maintaining high fidelity. Unlike the ideal case, in a dissipative environment the average polariton excitation number for $n$CMPB is necessarily lower than $n$. Nevertheless, as shown in Figs.~\ref{fig:figS7}\subref{fig:figS7b} and~\ref{fig:figS7}\subref{fig:figS7d}, the average excitation number can still remain at a relatively high level in the high‑fidelity region. For example, in Fig.~\ref{fig:figS7}\subref{fig:figS7b}, within the region with $F_p(4)>0.95$, the average excitation number lies roughly between 2.5 and 3.0, indicating that at least half of the population resides in the $|3\rangle$ and $|4\rangle$ states. Figure~\ref{fig:figS7} thus provides a useful reference for choosing the optimal drive strength parameters in a dissipative environment.
\vspace{8mm}

\twocolumn
\bibliographystyle{quantum}
\bibliography{citation}

\end{document}